\documentclass{emulateapj}
\usepackage{amssymb}
\usepackage{lscape}
\usepackage{longtable}
\begin{document}
\newcommand{\ha}{H$\alpha$}
\newcommand{\Ha}{\,{\rm H\alpha}}
\newcommand{\lya}{Ly$\alpha$}
\newcommand{\lyb}{Ly$\beta$}
\newcommand{\za}{$z_{\rm abs}$}
\newcommand{\ze}{$z_{\rm em}$}
\newcommand{\cmtwo}{cm$^{-2}$}
\newcommand{\nhi}{$N$(H$^0$)}
\newcommand{\degpoint}{\mbox{$^\circ\mskip-7.0mu.\,$}}
\newcommand{\kms}{\,km~s$^{-1}$}      
\newcommand{\minpoint}{\mbox{$'\mskip-4.7mu.\mskip0.8mu$}}
\newcommand{\peryr}{\mbox{$\>\rm yr^{-1}$}}
\newcommand{\secpoint}{\mbox{$''\mskip-7.6mu.\,$}}
\newcommand{\sqdeg}{\mbox{${\rm deg}^2$}}
\newcommand{\squig}{\sim\!\!}
\newcommand{\subsun}{\mbox{$_{\odot}$}}
\newcommand{\et}{{\rm et al.}~}
\newcommand{\msun}{\,{\rm M_\odot}}
\def\ltsima{$\; \buildrel < \over \sim \;$}
\def\simlt{\lower.5ex\hbox{\ltsima}}
\def\gtsima{$\; \buildrel > \over \sim \;$}
\def\simgt{\lower.5ex\hbox{\gtsima}}
\def\arcs{$''~$}
\def\arcm{$'~$}
\def\erf{\mathop{\rm erf}}
\def\erfc{\mathop{\rm erfc}}

\title{UV TO MID-IR OBSERVATIONS OF STAR-FORMING GALAXIES AT 
${\rm z \sim 2}$\altaffilmark{1}: STELLAR MASSES AND STELLAR POPULATIONS} 

\slugcomment{DRAFT: \today}
\author{\sc Alice E. Shapley\altaffilmark{2}}
\affil{University of California, Berkeley, Department of Astronomy, 601 Campbell Hall,
Berkeley, CA 94720}
\author{\sc Charles C. Steidel, Dawn K. Erb, and Naveen A. Reddy}
\affil{Calfornia Institute of Technology, MS 105-24, Pasadena, CA 91125}
\author{\sc Kurt L. Adelberger\altaffilmark{3}}
\affil{Observatories of the Carnegie Institution of Washington, 813 Santa Barbara Street,
Pasadena, CA 91101} 
\author{\sc Max Pettini}
\affil{Institute of Astronomy, Madingley Road, Cambridge CB3 OHA UK}
\author{\sc Pauline Barmby and Jiasheng Huang}
\affil{Harvard-Smithsonian Center for Astrophysics, 60 Garden Street, 
Cambridge, MA 02138}

\altaffiltext{1}{Based, in part, on data obtained at the 
W.M. Keck Observatory, which 
is operated as a scientific partnership among the California Institute of 
Technology, the
University of California, and NASA, and was made possible by the generous 
financial
support of the W.M. Keck Foundation. 
%Based in part on data obtained during the in-orbit
%check-out of the Spitzer Space Telescope.
}
\altaffiltext{2}{Miller Fellow}
\altaffiltext{3}{Carnegie Fellow}

\begin{abstract}
We present the broad-band UV through mid-infrared
spectral energy distributions (SEDs) of a sample of 72
spectroscopically-confirmed star-forming galaxies at
$z=2.30\pm 0.3$. Located in a $72$~arcmin$^2$ field
centered on the bright background QSO, HS1700+643, these
galaxies were pre-selected to lie at $z\sim 2$ based solely on
their rest-frame UV colors and luminosities, and
should be representative of UV-selected samples at high
redshift. In addition to deep ground-based photometry
spanning from $0.35 - 2.15 \mbox{ } \mu \mbox{m}$, we
make use of {\it Spitzer}/IRAC data, which probes the
rest-frame near-IR at $z\sim 2$.
The range of stellar populations present in the sample
is investigated with simple, single-component stellar 
population synthesis models. The inability to
constrain the form of the star-formation history limits
our ability to determine the parameters of extinction, age,
and star-formation rate without using external
multi-wavelength information. Emphasizing stellar mass
estimates, which are much less affected by these
uncertainties, we find $\langle log {M_{\ast} \over
M_{\sun}} \rangle = 10.32 \pm 0.51$ for the sample. The
addition of {\it Spitzer}/IRAC data as a long-wavelength
baseline reduces stellar mass
uncertainties by a factor of $1.5-2$ relative to
estimates based on optical -- $K_s$ photometry alone.
However, the total stellar mass estimated for the sample is
remarkably insensitive to the inclusion of IRAC
data. We find correlations
between stellar mass and rest-frame R band (observed
$K_s$) and rest-frame 1.4~$\mu$m (observed 4.5~$\mu$m)
luminosities, though with significant scatter. Even at
rest-frame $1.4 \mu$m, the $M/L$ ratio varies by a factor of 15
indicating that even the rest-frame near-IR, when taken alone, 
is a poor indicator of stellar mass in star-forming galaxies at $z\sim 2$.
Allowing for the possibility of episodic star formation, 
we find that typical galaxies in our sample could contain up
to three times more stellar mass in an old underlying
burst than what was inferred from single-component
modeling. In contrast, mass estimates for the most
massive galaxies in the sample ($M_* >
10^{11}M_{\odot}$) are fairly insensitive to the method
used to model the stellar population. Galaxies in this massive tail,
which are also the oldest objects in the sample,
could plausibly evolve into the passive galaxies
discovered at $z\sim 1.7$ with near-IR selection
techniques. In the general framework of hierarchical
galaxy formation and mergers, which implies
episodic star-formation histories, galaxies at high redshift may
pass in and out of UV-selected and near-IR
color-selected samples as they evolve from phases of
active star formation to quiescence and back again.

\end{abstract}
\keywords{cosmology: observations --- galaxies: evolution --- galaxies: high-redshift --- 
galaxies: starburst --- stars: formation}

\section{Introduction}
\label{sec:intro}

Within the last few years, new techniques and facilities
have resulted in a burst of activity focused on
understanding the progress of galaxy formation during the
era near $z \simeq 2$. While this epoch has been known
for some time as the peak of the ``quasar era''
(\citet{schmidt1995}), until recently it has comprised
part of the ``redshift desert'' about which little was
known for normal galaxies.  The recent exploration of
galaxies in the $z \sim 2$ era has occurred on a number
of fronts:  new surveys have now successfully identified
samples of $z \sim 2$ galaxies based upon their very
bright far-IR and radio emission 
\citep[e.g,][]{chapman2004}, their signature rest-frame far--UV
colors \citep{adelberger2004a,steidel2004}, their high
luminosity in the observed $K$ band \citep{daddi2004},
red rest-frame optical colors
\citep{vandokkum2003,franx2003}, or some combination of the
above \citep[e.g.,][]{abraham2004,shapley2004}.  
The results of most of these surveys suggest that it was
near $z \sim 2$ that galaxies as massive as any we see in
the local universe began reaching maturity, and where a
transition from rapidly star-forming to asymptotically
quiescent may be occurring ``before our eyes'' for
relatively massive galaxies \citep[e.g.,][]{mccarthy2004,
daddi2004b,shapley2004}. Fortuitously,
galaxies in the redshift range $2 \simlt z \simlt 2.5$
are particularly well-suited to spectroscopic studies
both in the observed optical \citep{steidel2004,daddi2004}
and especially in the observed near-IR, where
nebular lines valuable for measurement of kinematics,
star-formation rates, and chemical abundances are
observationally accessible \citep[e.g.,][]{erb2003,
erb2004,erb2005,shapley2004,vandokkum2004}.

One of the fundamental physical measurements that can be
made for high redshift galaxies, and one of the more
important for placing them into the theoretical context,
is the stellar mass. A number of authors in the last
several years have interpreted observed-frame UV to near-IR
photometry with stellar population synthesis models
in order to infer the stellar mass for samples of $z
\simgt 2$ galaxies \citep[e.g.,][]{papovich2001,
shapley2001,daddi2004}. Others have used
similar methods and the technique of photometric
redshifts to obtain measures of the total stellar mass as
a function of redshift \citep[e.g.,][]{dickinson2003,
fontana2004,glazebrook2004}. 
Stellar mass measures become
increasingly uncertain as one goes to higher redshift,
where ground-based observations, limited to $\lambda
\simlt 2.3$ $\mu$m, probe increasingly shorter rest-frame
wavelengths. At $z>2$, the $K_s$ band corresponds to the rest-frame
optical, where current or recent star formation may dominate 
the observed luminosity rather than
the stellar mass that has accumulated over a galaxy's
lifetime. For this reason, the arrival of
data from the IRAC instrument \citep{fazio2004} on board
the {\it Spitzer Space Telescope} (hereafter {\it Spitzer}) has
been widely anticipated as a means to significantly
improve the measurement of stellar mass and star
formation history since it is capable of obtaining very
sensitive observations out to 8$\mu$m, allowing
photometric access to the rest--frame $1-2\mu$m light
from galaxies to redshifts $z \sim 4$, and providing
particularly sensitive observations of this region for
galaxies with more modest redshifts near $z\sim 2$.

In this paper, we discuss the broad-band properties of a
reasonably large sample of spectroscopically confirmed,
UV-selected galaxies at $z \sim 2$ in a field that was
observed by the IRAC ``In-Orbit Checkout'' (IOC) program 
in the early days of {\it Spitzer}. The principal aim has been to 
explore the broad-band
spectral energy distributions (SEDs) for a robustly
defined sample of high redshift galaxies that naturally
spans a large range in stellar mass, star-formation rate,
and star-formation history. The main advantages of a
UV-selected sample at $z \sim 2$, described in detail
elsewhere \citep{adelberger2004a,steidel2004} are
that confirming optical spectra can be easily obtained,
photometric selection is highly efficient and requires
using only relatively ``inexpensive'' optical
ground-based photometry, and it is possible to tune color criteria 
to yield a roughly volume-limited sample over the desired 
redshift range.
In this particular case, an additional advantage was that
the spectroscopic sample of galaxies already existed, as
did deep photometry from observed 0.35 to 2.15 $\mu$m and
targeted near-IR spectroscopic observations. Details of
the near-IR photometry and spectroscopy will be presented
in \citet{erb2005}. The HS1700+6416 field is one of several
we have used in a survey designed to combine observations
of $z \sim 2$ galaxies with sensitive observations of the
intergalactic medium (IGM) in the same volume. In this
respect, HS1700+6416 ($z_{em}=2.72$) is one of the very
best-studied high redshift QSOs in the entire sky, due to
its unusual brightness ($V \sim 16$). It also happens to
lie in the ``continuous viewing zone'' for both {\it Spitzer} and
HST; this fact, plus the availability of the ancillary
deep ground based data, is the main reason the field was
chosen for the deep IOC observations by the IRAC team.

In combination with the ground-based photometric and
spectroscopic data, the IRAC observations of this field
provide a means to establish the range of properties
represented among galaxies selected using UV colors;
these may be useful for comparison to galaxy samples at
similar redshifts but selected using other techniques.  
Of particular interest to us is a quantitative assessment
of the ``value added'' by longer wavelength photometry
for assessing the galaxy properties at high redshifts as
compared to what one would infer from ground-based
measurements alone (since realistically only a fraction
of ground-based fields will be observed by {\it Spitzer} during
its lifetime). As it turns out, the HS1700 field is also
interesting because it allows for a preliminary
comparison of the dependence of $z \sim 2$ galaxy
properties on large-scale environment
\citep{steidel2005}.

The paper is organized as follows: \S\ref{sec:obs}
presents the ground-based and {\it Spitzer} observations;
\S\ref{sec:phot} describes the photometry and the
selection of the sample which is discussed in the rest of
the paper;  \S\ref{sec:mod} describes the use of stellar
population synthesis models to infer physical properties
of the galaxies from their broad-band spectral energy
distributions (SEDs); \S\ref{sec:results} summarizes the
results of the modeling focusing on the galaxy stellar masses and
mass-to-light ratios $(M/L)$; 
\S\ref{sec:discussion} discusses the results
and their general implications; \S\ref{sec:summary}
summarizes the main results.

We assume a cosmology with $\Omega_m=0.3$,
$\Omega_{\Lambda}=0.7$, and $h=0.7$ throughout.

\section{Observations}
\label{sec:obs}

\subsection{Optical Imaging}
\label{sec:obsopt}

Images of the HS1700$+$64 field were obtained in 2001 May
using the William Herschel 4.2m telescope on La Palma and
the Prime Focus Imager. The WHT system uses a 2-chip
mosaic of Marconi 2k x 4k CCDs with especially good UV
quantum efficiency, and provides a field of view of
16\arcm\ by 16\arcm\ sampled at 0\secpoint236 per pixel.
The conditions during the observing run were excellent,
with sub-arcsecond seeing in all bands. The field was
roughly centered on the position of the QSO, and 15
arcsecond dithers were executed between subsequent
exposures in order to allow for fringe removal in the
${\cal R}$ band and the creation of dark sky superflats
(particularly necessary in the $U_n$ band).  The data
were reduced using procedures described in detail by
\citet{steidel2003}.

Additional images in the $U_n$ band covering the central
5\arcm\ by 7\arcm\ region of the WHT field were obtained
in 2001 May using the blue side of the Low Resolution
Imaging Spectrometer on the Keck I 10m telescope
\citep{mccarthy1998,steidel2004}. These were scaled,
variance weighted and combined with the WHT data to form
the final $U_n$ image.

Properties of the reduced and stacked images are
summarized in Table~\ref{tab:obs}.

\subsection{Near-IR Imaging}
\label{sec:obsnearir}

We observed the HS1700$+$64 field during two observing
runs, 2003 May and 2003 October, using the Palomar 5.1m
Hale telescope and the Wide Field Infrared Camera
\citep[WIRC;][]{wilson2003}. The images were taken in the
K-short ($K_s$) filter, obtained in a sequence of
30-second integrations, moving the telescope in a
prescribed dither pattern after each sequence of 4
exposures. The conditions during both observing runs were
generally good, with seeing ranging between 0\secpoint55 
and 0\secpoint99, but the
ambient temperature was warm resulting in relatively high
$K_s$ band sky backgrounds ($\sim 12.4$ mag
arcsec$^{-2}$).  The WIRC camera employs a Rockwell
HgCdTe ``Hawaii-2'' 2k x 2k array, providing a field of
view of 8\minpoint5 x 8\minpoint5 with a spatial sampling
of 0.249 arcsec per pixel.

The WIRC data from each (typically 54 minute) dither
sequence were reduced, registered, and stacked using a
set of IDL scripts written by K. Bundy (private
communication). The resulting images were then registered
and combined using IRAF tasks. The final $K_s$ image
represents a total of 11.03 hours of integration in the
parts of the image that received the full integration
time. The final mosaic was trimmed to include only the
parts of the image receiving 50\% or more of the total
integration time ($\sim 90\%$ of the trimmed image
received the full exposure time).  The image was
calibrated with reference to the near-IR standards of
\citet{persson1998}.

\subsection{{\it Spitzer}/IRAC Imaging} 
\label{sec:obsmidir}

Observations with the Infrared Array Camera (IRAC) were
obtained in 2003 October during the ``In-Orbit Checkout'' (IOC)
of {\it Spitzer}. The observing strategy and data
reductions are described in detail by \citet{barmby2004}.
The IRAC observations were designed to overlap as much as
possible with the ground-based $K_s$ and optical images,
but some compromises were adopted to avoid known
foreground galaxy clusters in the field, so that the IRAC
pointings were biased toward the SW relative to the
ground-based data, which are centered on the bright QSO.
As detailed below, the IRAC images in the 4 IRAC bands
(see Table~\ref{tab:obs}) completely overlap over only a
relatively small fraction of the field observed in the
$K_s$ band described above, with the smallest overlap in
the IRAC channel 1 (3.6$\mu$m) and channel 3 (5.8 $\mu$m)
images, but much larger overlap in channels 2 (4.5$\mu$m)
and 4 (8.0$\mu$m).  Figure ~\ref{fig:map} illustrates the
regions covered in the IRAC images compared to those
covered in the Palomar $K_s$ image.

The photometric calibrations for the IRAC images were as
determined by the IRAC team \citep{barmby2004}; however,
we used a significantly different method for performing
the photometry and estimating the photometric uncertainty
(see ~\ref{sec:photirac}), resulting in significantly
more conservative (i.e. larger) estimated errors at a
given flux density.

\begin{figure*}
%\plotone{map.eps}
%\plotone{f1.eps}
\plotone{f1_blank.eps}
\caption{A schematic map of the survey field considered.
The grayscale image is the Palomar 5m WIRC $K_s$ image,
with the positions of objects with $1.4 \le z \le 2.8$
marked with black circles. The colored contours (blue=IRAC/Ch~1;
green=IRAC/Ch~2; yellow=IRAC/Ch~3; red=IRAC/Ch~4) indicate
regions of the image that received $>10$\% of the total
IRAC integrations time in each IRAC band, i.e., $>1.1$
hours. For full resolution figure, go to http://astron.berkeley.edu/$\sim$~aes.}
\label{fig:map}
\end{figure*}

\section{Photometry}
\label{sec:phot}

\subsection{Optical and Near-IR}
\label{sec:photoptir}

The observables from our ground-based images are $\cal R$ magnitudes
and isophotal ${\cal R}-K_s$, $G-{\cal R}$, and $U_n-G$ colors,
with isophotes defined in the $\cal R$ detection catalog. 
In order to mimic the actual observing process,
we therefore estimated uncertainties in isophotal
color rather than magnitude for all
optical/IR bands except ${\cal R}$. The uncertainties
were estimated separately for the optical and IR
photometry, by adding large numbers of simulated galaxies
of known magnitudes to the images and comparing their
recovered photometry with their input parameters. The
simulated galaxies were drawn from templates of
exponential disks with scale lengths ranging from
0\farcs05 to 0\farcs4, smoothed to match the seeing of
the individual images.  For the ${\cal R}-K_s$
uncertainties, we selected galaxies at random within the
range $21.0 < {\cal R} < 26.0$ and $1.0 <{\cal R}-K_s <
5.5$ and placed them at random positions on the $\cal R$
and $K_s$ images, 200 at a time (to avoid overcrowding) for
a total of 250,000 fake galaxies.  We then performed
photometry as usual on the images, and compared the
catalog of detections with the input list of simulated
galaxies.  We considered the quantity $\Delta [{\cal
R}-K_s]= ({\cal R}-K_s)_{meas}-({\cal R}-K_s)_{true}$, and
binned in steps of 0.5 mag in ${\cal R}_{meas}$ and 0.2
mag in $({\cal R}-K_s)_{meas}$.  The mean value of $\Delta
[{\cal R}-K_s]$ is an estimate of the bias in the
photometry; for typical objects, $\Delta [{\cal R}-K_s] <
0.05$ mag.  The standard deviation $\sigma (\Delta [{\cal
R}-K_s])$ provides an estimate of the ${\cal R}-K_s$
uncertainty for objects that fall in each bin; this was
typically $\sigma (\Delta [{\cal R}-K_s]) \sim 0.25$ mag.  
${\cal R}-K_s$ uncertainties for each object are given in
Table~\ref{tab:phot}.  Uncertainties in ${\cal R}$,
$G-{\cal R}$ and $U_n-G$ were determined in a similar
fashion, by adding large numbers of fake galaxies to the
$U_n$, $G$ and ${\cal R}$ images and comparing their
recovered and input properties, with the additional
restriction that only those objects whose recovered
photometry met our selection criteria were considered. We
binned in ${\cal R}$ and $G-{\cal R}$ to determine the
${\cal R}$ and $G-{\cal R}$ uncertainties, which were
typically $\sigma (\Delta {\cal R}) \sim 0.14$ mag and
$\sigma (\Delta [G-{\cal R}]) \sim 0.07$ mag, and in $G$
and $U_n-G$ to determine $U_n-G$ uncertainties, typically
$\sigma (\Delta [U_n-G]) \sim 0.13$ mag.  Again
uncertainties for each object are given in
Table~\ref{tab:phot}. Similar techniques for estimating
photometric uncertainties have been used in, e.g.,
\citet{adelberger2000,steidel2003,shapley2001}.

\subsection{IRAC}
\label{sec:photirac}

We measured IRAC fluxes by first computing an empirical
point spread function (PSF) for each channel by median
averaging the signal from several isolated point sources
located throughout the mosaiced IRAC data in the HS1700
field and normalizing the resulting PSF to have unit
flux.  We obtained accurate object positions ($\simlt
0.2$\arcs\ rms uncertainty) for $>$95\% of IRAC sources
in the field from their $K_s$-band counterparts in
the WIRC imaging of this field, or from their
${\cal R}$-band counterparts for the small fraction of 
objects detected in ${\cal R}$ but not in $K_s$. 
We then used the empirically-determined IRAC PSF to simultaneously
model the emission of all objects within a sub-image of
size 24\arcs\ surrounding each object of interest,
similar to methods generally used for photometry of
stellar objects in crowded fields.  This procedure to
some extent mitigates the effects of confusion noise,
since the object positions are known {\it a priori} from
the much higher spatial resolution $K_s$ band image, and
allows for some de-blending of objects that are only
partially resolved in the IRAC images.  The method is
also similar to that employed by \citet{fs1999} and
\citet{papovich2001} for comparing ground-based near-IR
and HST/WFPC-2 photometry.  Photometric errors were
computed by adding in quadrature the error in the source
flux (Poisson noise) and the dispersion in measured flux
values found by fitting PSFs to 100 random positions
containing no obvious sources near the object of
interest. This latter estimate should account to a large
extent for noise due to source confusion, which is
particularly relevant in the 3.6 and 4.5 $\mu$m bands.
Except in cases of very bright sources, the background
dispersion dominates the errors.  The PSF full-width at
half maximum (FWHM) and approximate 5$\sigma$ sensitivity
limits for each channel are presented in
Table~\ref{tab:phot}.  We omitted any measured
fluxes with uncertainties larger than 0.5 magnitudes from
consideration in fitting the SEDs as described in
\S~\ref{sec:modsamp}.  In addition, we
set the errors to 0.10 magnitudes for measurements whose
formal uncertainties calculated using the above
prescription were $< 0.10$ magnitudes, in order to
account for systematic uncertainties in the photometry
that are not otherwise accounted for (e.g., source
confusion on scales smaller than the PSF, estimation of
local background, etc.). 

\begin{figure}
%\plotone{zhist_72fit.eps}
\plotone{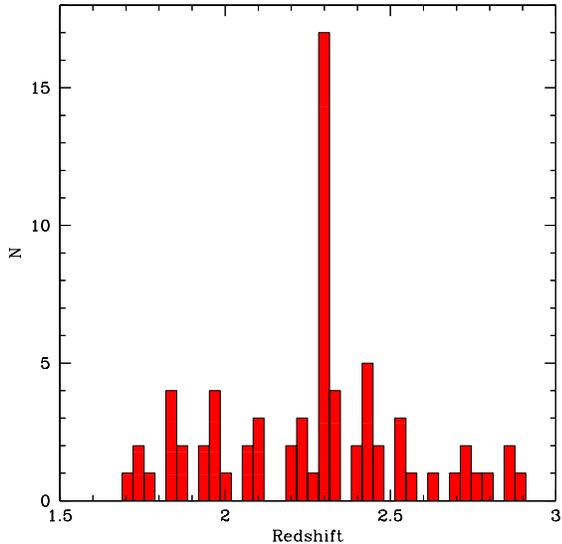}
\caption{
Redshift histogram for 72 spectroscopically identified
galaxies in the HS1700$+$643 field with near-IR and
IRAC detections. This histogram contains
a significant spike at $z=2.30\pm0.015$, discussed by
\citet{steidel2005}.
}
\label{fig:zhist}
\end{figure}

\subsection{Sample Selection}
\label{sec:sample}

Because the redshift of HS1700+643 is $z_{em} = 2.72$ and
we were interested in the properties of the galaxies in
the ``foreground,'' the spectroscopic observations
targeted photometrically selected galaxies that would
neatly cover the redshift range $1.9 \simlt z \simlt
2.7$. To accomplish this we selected a combination of
objects satisfying either the ``BX'' photometric criteria
of \citet{adelberger2004a} ($z=2.20\pm0.32$) or the ``MD''
photometric criteria of \citet{steidel2003}
($z=2.79\pm0.27)$. A total of 100 objects (81 ``BX'' and
19 ``MD'') has measured spectroscopic redshifts $z > 1.4$
from observations with the Low Resolution Imaging
Spectrometer, described in detail in \citet{steidel2004}.  
Because the surface density of BX candidates is $\sim 6$ times 
higher than that of MD candidates to the same ${\cal R}=25.5$
apparent magnitude limit, the expected redshift
distribution for a BX/MD photometric sample, based on the total of 1133
spectroscopic redshifts for BX/MD candidates in all
survey fields, is $\langle z \rangle = 2.24\pm0.36$. 

The sample of galaxies discussed in the rest of the paper
is culled from the region of overlap between the
ground-based $U_nG{\cal R}$ images, the ground-based
$K_s$ image, the 4 IRAC bands (see figure~\ref{fig:map})
and the spectroscopic observations of UV-selected
galaxies. Within the 8\minpoint3 by 7\minpoint8 region
covered by the Palomar $K_s$ image, there is a total of
508 BX and MD-type galaxy candidates identified from the
optical photometry using the selection criteria defined
by \citet{adelberger2004a} and \citet{steidel2003},
respectively; of those, 389 (77\%) of the objects are
significantly detected in the $K_s$-band image. 
Within the 64.7 square arcmin region covered by the $K_s$ image,
we obtained spectra for 162 of the 508 photometric candidates
and spectroscopically
identified 98 objects. The identified fraction represents a 60.5\%
spectroscopic success rate, similar to
that of our $z\sim 2$ survey overall \citep{steidel2004}.
As discussed in \citet{steidel2004}, we expect the
unidentifed objects have the same redshift distribution
as the spectroscopic successes. Of the 98 objects with spectroscopic
identifications, 92 are detected in the $K_s$ image
to $K_s(\rm Vega)\sim 22$; of the 92 objects with
spectroscopic IDs and $K_s$ detections, 79 (86\%) are at
$z > 1.4$ and have a mean redshift and dispersion of
$2.26\pm 0.30$, entirely consistent with the expected
redshift distribution for all BX/MD candidates as
discussed above. Of the 79 spectroscopically identified
$z \sim 2$ galaxies, 72 are detected in one or more of
the IRAC passbands and included in the stellar population
analysis. Invariably, objects detected
significantly at 2.15 $\mu$m are also detected in the
IRAC 3.6 and 4.5 $\mu$m bands; the 7 galaxies detected at
2.15$\mu$m but not with IRAC simply fall outside of the
well-exposed regions of the IRAC pointings.

In summary, the 72 $z \sim 2$ galaxies discussed in this
paper are representative of a UV-selected sample to
${\cal R}=25.5$, but are objects which are also detected
in both the near and mid-IR. The only possible bias of
this sample relative to any sample selected using the
same UV criteria is that it excludes the $\sim 23$\% of
BX/MD candidates in the field that are not detected to
$K_s \sim 22$. The redshift distribution for this
spectroscopic sample with near and mid-IR detections
is shown in figure~\ref{fig:zhist}. Very strikingly,
there is a significant galaxy overdensity in this
distribution at $z=2.30$, which allows for an examination of
galaxy properties as a function of large-scale environment
\citep{steidel2005}.

\begin{figure*}
%\plotone{iracugr2.eps}
\plotone{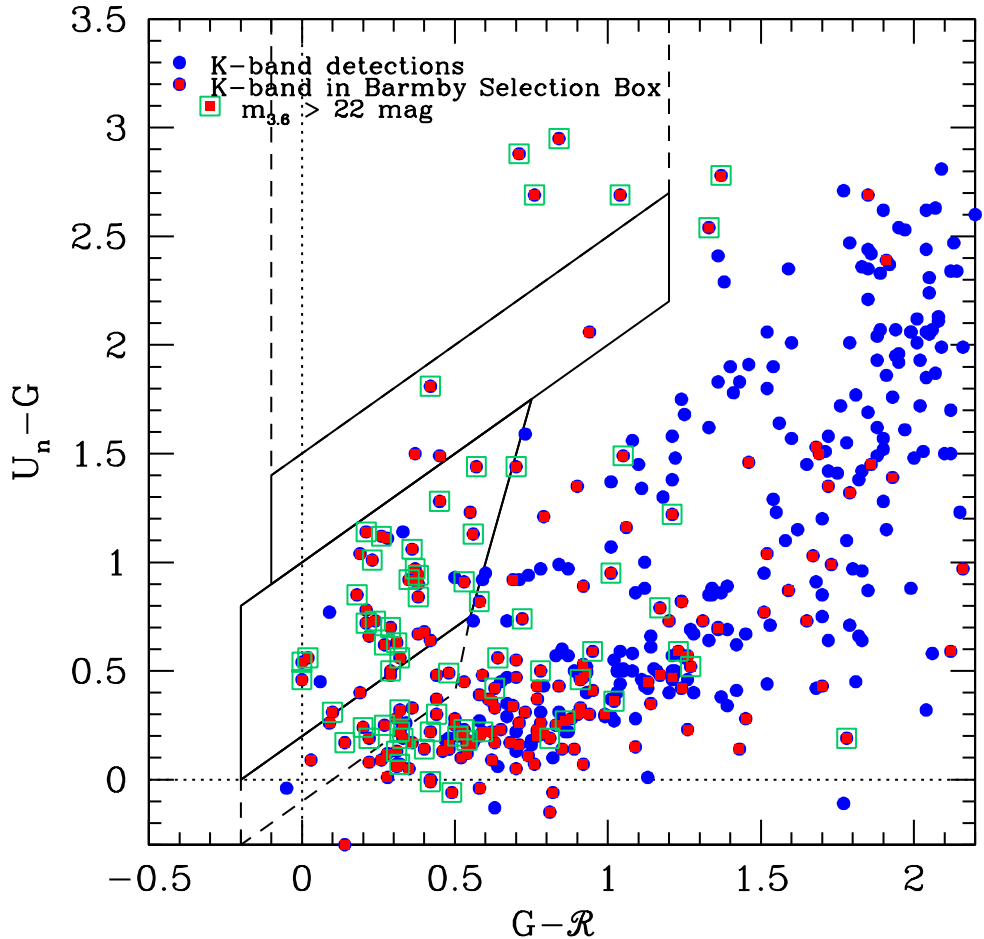}
\caption{
Plot showing the location in optical $U_nG{\cal R}$ color
space of all objects in a K-selected catalog to
$K\simeq22$. The regions within the solid black lines
denote those used to select the galaxies discussed in
this paper (``BX'' and ``MD''); the dotted curves denote
regions containing ``BM'' ($z=1.7\pm0.3$, lower left) and
``C/D'' LBGs ($z=3.0\pm0.3$, upper left). Objects
satisfying the ``high z'' galaxy criteria suggested by
Barmby \et (2004)  based on IRAC colors are marked in
red. As discussed in the text, a significant fraction
($\sim 40$\%) of the IRAC ``high z'' objects in the
optical catalog appear to lie at $z < 1.4$.}

\label{fig:iracugr}
\end{figure*}

It is of interest, for some purposes, to ask what
fraction of a mid-IR selected galaxy sample at the same
redshift is represented by objects selected using the UV
color criteria. This is a difficult question to answer,
as there is as yet no large spectroscopic sample of
mid-IR selected high redshift galaxies (and, as we argue
below, photometric redshifts may be an inadequate
approach to the problem).  However, we can make a rough
estimate as follows. Recently, \citet{barmby2004}
suggested a region in IRAC 3.6/4.5/5.8 $\mu$m color space
that may make it possible to select high redshift
galaxies from IRAC colors alone, based on models and the
observed locations of $z \sim 3$ UV-selected galaxies in
IRAC color space.  In figure~\ref{fig:iracugr} we show
the locations in the $U_nG{\cal R}$ color plane of all of
the objects detected in the $K_s$ band that are within
the region of the IRAC images covered by IRAC at 3.6,
4.5, and 5.8$\mu$m and which have IRAC uncertainties of
$<0.3$ mag in each band.\footnote{The size of this region
is approximately 30 arcmin$^{2}$, although only $\sim
40$\% of the objects in a $K_s$-selected catalog within this
region are included, primarily because of the limited
depth of the 5.8$\mu$m image.}. Out of a total of 618
galaxies with $K_s$, 3.6$\mu$m, 4.5$\mu$m, and 5.8$\mu$m
measurements, 443 are included in the optical photometric
catalog to ${\cal R}=25.5$, 99 of which satisfy the
criteria proposed by ~\cite{barmby2004} in the
3.6-5.8$\mu$m color space.  As shown in figure~\ref{fig:iracugr}, 
46 of the 99 sources satsifying the
\citeauthor{barmby2004} color criteria do indeed lie in the regions of
$U_nG{\cal R}$ color space from which $z \sim 2$ BM
($z=1.70\pm0.34$), BX ($z=2.20\pm0.32$), MD
($z=2.72\pm0.3$) and C/D/M ($z=3.0\pm0.3$)
\citep{steidel2003} galaxies are drawn. However, there is
a significant number of galaxies ($\sim 40$) which
satisfy the Barmby \et criteria but are located in
regions inhabited primarily by $z \sim 0.7-1.5$ star
forming or passive galaxies in optical color space
\citep[see, e.g.,][]{adelberger2004a}. Another $\sim 10$ have
optical colors that are outside of the UV selection
windows, but are consistent with more reddened star
forming galaxies at $z \sim 1.5-2.5$ \citep[cf.][]{daddi2004b}.
Of the 175 galaxies that are well detected in the
$K_s$, 3.6, 4.5, and 5.8 $\mu$m bands but absent from the
optical catalog, 81 satisfy the \citeauthor{barmby2004} color
criteria. Assuming that the same percentage ($\sim 40$\%)
of these are likely to be in the foreground as for the
optically detected galaxies, $\sim 50$ galaxies to $K_s
\sim 22$ are plausibly at $z \simgt 1.5$\footnote{29 of
the 81 objects are actually detected in the ${\cal R}$
band to ${\cal R}\sim 26$ but are not included in
figure~\ref{fig:iracugr} because the colors are not
reliably measured.}. Most of the galaxies in this red
sample are relatively faint in the $K_s$ band, with only 12
of 81 having $K_s < 20$, of which perhaps 6-8 are high
redshift objects.

In summary, of the 180 galaxies well detected in the
in the $K_s$, 3.6, 4.5, and 5.8 $\mu$m bands that
also satisfy the \citeauthor{barmby2004} criteria,
46 would be selected based on $U_nG{\cal R}$ colors to
be at $z \simgt 1.5$, $\sim 70$ ($\sim 40$\%) are likely to be
at $z\simlt 1.5$, while the remaining $\sim 60$ galaxies
are plausible $z\simgt 1.5$ candidates that do not
satisfy the $U_nG{\cal R}$ selection criteria.
Thus, we conclude that $\sim 50$\% of galaxies at $z
\sim 2.3\pm0.4$ to an apparent magnitude limit of $K\sim
22$ ($K\sim 23.8$ on the AB system) would be selected
using the rest-UV color criteria to ${\cal R}=25.5$.  
This statistic is broadly consistent with the results of
the FIRES survey \citep{franx2003} which targeted
galaxies with colors $J-K>2.3$, most of which are
similarly optically faint, as well as with the ``IRAC
EROs'' discussed by \citet{yan2004}.  We defer further
discussion of the overlap and distinctiveness of samples
selected in various ways to future work.

\section{Population Synthesis Modeling}
\label{sec:mod}

\subsection{Sample}
\label{sec:modsamp}

In order to explore the stellar populations of
UV-selected $z\sim 2$ galaxies in the HS1700+64 IRAC
pointing, we considered only the subsample of 72 galaxies
with measured redshifts, $K_s$ magnitudes, and IRAC
detections in at least one channel. As discussed below,
the set of colors and magnitudes for BX and MD galaxies
is not sufficient to simultaneously determine accurate
photometric redshifts and stellar masses for galaxies
that do not have spectroscopic redshifts.

The photometric measurements used to model the stellar
populations are given in Table 2.  Note that all of the
magnitudes and colors are on the AB system, with the
exception of the ${\cal R}-K_s$ color, for which the
$K_s$ mag is on the Vega system for easier comparison to
other work on similar high redshift
galaxies.\footnote{$K_s$ magnitudes or ${\cal R}-K_s$
colors may be converted to the AB system by adding 1.82
to the tabulated values.} For the 13 galaxies at $z>2.5$,
we did not include the $U_n-G$ color in the SED fit,
since absorption by H~I in the intergalactic medium
begins to have a strong effect on the measured $U_n$
magnitude at these redshifts.

Before modeling the galaxy magnitudes and colors, we
corrected the $U_n$ or $G$ magnitude for the effects of
galactic Ly$\alpha$ emission and absorption. Ly$\alpha$
equivalent widths were measured from the LRIS spectra
used to identify redshifts. These corrections were
applied to galaxies at $z=1.7-1.9$ ($U_n$) or $z>2.47$
($G$). We also have $\Ha$ and [NII] emission line
measurements for four galaxies in the HS1700 sample
(BX691, BX717, MD103, and MD109, all from
\citet{erb2003}). We corrected the observed $K_s$ by
0.10, 0.17, 0.08, and 0.10 magnitudes, respectively to
account for the contribution of nebular emission to the
observed flux. $K_s$ magnitudes for the remainder of the
galaxies in the redshift range $2.016 < z < 2.504$ remain
uncorrected.  Based on a much larger sample of H$\alpha$
spectra of galaxies in the same redshift range (which
will be discussed in detail by \citet{erb2005}), the
typical correction to the broad-band $K_s$ flux is
$+0.09\pm0.07$ magnitudes, although some of the most
extreme cases have H$\alpha$ equivalent widths large
enough to change the $K_s$ magnitude by as much as
several tenths of a mag.  The possible impact of the
$K_s$ corrections on the modeling results is discussed
below.

Only 22\% of the BX/MD photometric candidates with both
$K_s$ photometry and measurements in at least one IRAC
band are spectroscopically identified, and we restrict
the stellar population modeling to this spectroscopic
sample. To test the potential success of modeling
photometric candidates without redshifts, we tried to fit
simultaneously the redshift and stellar population
parameters for the sample of galaxies with spectroscopic
redshifts, pretending that the redshift was an
undetermined free parameter.  With only $U_nG{\cal R}K_s$
plus IRAC photometry, we obtained an accuracy of $\langle
| z_{phot} - z_{spec} |/ (1+z_{spec}) \rangle = 0.09$. At
the mean redshift of the sample, this error corresponds
to a typical $|z_{phot} - z_{spec}|$ of 0.3, while in
several cases the redshift estimates fail
catastrophically, with $|z_{phot} - z_{spec}| > 1$.  We
note that this accuracy is very similar to that found for
photometric redshift estimates in the FIRES survey, based
on 39 galaxies with spectroscopic redshifts in the {\it
Hubble Deep Field South} (HDF-S) \citep{rudnick2003,
labbe2003}. Perhaps more significantly, the stellar
population parameters inferred when simultaneously
fitting for the redshift do not uniformly agree with
those found when the redshift is known.  Specifically, we
find a typical fractional stellar mass difference of
$\langle |M_{\ast}^{phot} - M_{\ast}^{spec}| /
M_{\ast}^{spec} \rangle = 0.56 \pm 1.23$, which is larger
than any of the fractional stellar mass uncertainties we
find due to photometric uncertainties, or uncertainties
in our choice of star-formation history for
single-component models (see discussion in
section~\ref{sec:resultsmass}). Not only is the scatter
in this fractional offset quite large, but the
distribution is skewed toward overestimating the mass,
such that the average $\langle (M_{\ast}^{phot} -
M_{\ast}^{spec})/M_{\ast}^{spec} \rangle = 0.37 \pm
1.30$.\footnote{ When we exclude five galaxies with
wildly discrepant $M_{\ast}^{phot}$ relative to
$M_{\ast}^{spec}$ (all are overestimates), we find a
relatively symmetric distribution around zero, with
$\langle (M_{\ast}^{phot} -
M_{\ast}^{spec})/M_{\ast}^{spec} \rangle = 0.07 \pm
0.44$.} This exercise demonstrates that, with only
$U_nG{\cal R}K_s$ plus IRAC photometry, even for objects
known to have redshifts in the targeted range (which
presumably reduces the probability of catastrophic
failures), considerable uncertainties would be introduced
if objects without spectroscopic redshifts were included.

\subsection{Modeling Procedure}
\label{sec:modproc}

Following a procedure similar to that of
\citet{shapley2001} and \citet{shapley2004}, we apply
\citet{bc2003} models to fit the SEDs 
of the UV-selected $z\sim 2$
galaxies in the HS1700 {\it Spitzer} pointing. We used models
with solar metallicity and a Salpeter initial mass
function (IMF) extending from
$0.1 - 100 M_{\odot}$. Also, as recommended by
\citet{bc2003}, we adopted the Padova 1994 stellar
evolution tracks. \citet{papovich2001} have
investigated how the best-fit parameters depend
systematically on the choice of metallicity and IMF,
though we will not include such a discussion
here. We do note that, at least
for galaxies bright in the rest-frame optical ($K_s \leq
20$), solar metallicity appears to be a good
approximation, and typical $z\sim 2$ galaxies appear to
have metallicities that are only slightly lower than
solar \citep{shapley2004}.
We also note that
it is well established that a Salpeter IMF
over-predicts the number of stars less massive than 1
M$_{\sun}$ (or, alternatively, that it over-predicts the
stellar $M/L$) compared to observations in the local
universe \citep[e.g.,][]{bell2003}. We use the Salpeter IMF
to facilitate comparison with other work 
\citep[e.g.,][]{cole2001,shapley2001}; use of the
\citet{chabrier2003} IMF, the IMF proposed by
\citet{bg2003}, or the ``diet'' Salpeter IMF used by
\citet{bell2003}, would reduce the inferred stellar masses
by a factor of $\sim 1.5-1.8$. 
Dust extinction is taken
into account with a \citet{calzetti2000} starburst
attenuation law. \citet{reddy2004} have shown that using
the Calzetti law and continuous
star-formation models to infer unobscured star-formation rates
from UV colors and magnitudes accurately predicts the
{\it average} X-ray and radio continuum fluxes of $z\sim 2$
star-forming galaxies, though the object to object scatter
may be large.
In order to
determine how well the stellar population parameters can
be constrained, we investigated a range of star-formation
histories of the form $SFR(t) \propto \exp(-t/\tau)$,
with e-folding times of $\tau=0.01, \mbox{ } 0.05, \mbox{
} 0.10, \mbox{ } 0.2, \mbox{ } 0.5, \mbox{ } 1, \mbox{ }
2, \mbox{ and } 5$~Gyr, as well as continuous
star-formation (CSF) models. For the main analysis, we
did not consider more complex star-formation histories,
such as short-duration bursts occurring at random past
times superimposed on a smoothly declining star-formation
history. The data for most galaxies do not place strong
constraints among the range of simple, single-component
star-formation histories considered, and therefore the
use of a more complex model does not seem justified.
We only briefly employ extreme two-component models in 
section~\ref{sec:resultsmcorr}, in order
to determine how much stellar mass could be contained
in these galaxies in maximally old underlying bursts.

Our goal was to constrain the stellar population
parameters of each galaxy. The fitted parameters 
included: dust extinction (parameterized by $E(B-V)$),
age ($t_{sf}$), star-formation rate (SFR), stellar mass
($M_{\ast}$), and star-formation history ($\tau$). 
For each $\tau$, we used a grid of \citet{bc2003} models with
ages ranging between 1 Myr and the age of the Universe at
the redshift of the galaxy being modeled and extinctions
ranging between $E(B-V)=0.0$ and $E(B-V)=0.7$ (though
none of the galaxies has a best-fit $E(B-V) > 0.5$).
For each combination of $E(B-V)$, $t_{sf}$, and $\tau$, we
computed the full $U_nG{\cal R}K_s$ plus
IRAC $3.6 - 8.0 \mu$m SED of the model placed at
the redshift of the observed galaxy.  In computing this
SED, we reddened the intrinsic model
galaxy spectrum by dust, and further attenuated it in a
manner simulating absorption by the intergalactic medium
of neutral hydrogen \citep{madau1995}, which affects the
predicted $U_n$ and $G$ magnitudes. \citet{bc2003}
CSF models are by default normalized to have
star-formation rates of $1 M_{\odot} \mbox{yr}^{-1}$, while
declining models are normalized to have a total mass of $1 M_{\odot}$
at $t=\infty$. We searched for
the model normalization that minimized the value of
$\chi^2$ with respect to the measured SED,
which yielded an estimate of the 
star-formation rate and stellar mass. 
The combination of $E(B-V)$, $t_{sf}$, $\tau$ and normalization
yielding the lowest $\chi^2$ was called the overall ``best-fit''.
Figure~\ref{fig:seds1} shows the overall best-fit model
together with the observed data points for each of the 72
galaxies in the sample.  In most cases the 
``best-fit'' $\tau$ model is not significantly better
than that of the best-fitting model for every other value
of $\tau$ considered, as discussed below.

\begin{figure*}
\epsscale{0.9}
%\plotone{mc_plots.eps}
%\plotone{f4.eps}
\plotone{f4_blank.eps}
\epsscale{1.0}
\caption{
Examples of confidence intervals obtained from Monte
Carlo modeling for three different galaxies selected to
span a range of inferred properties, as described in the
text (BX879, BX691, and BX563).
The confidence intervals have been projected onto the spaces of
(E(B-V),Age) and (Age,Mass), and the
dark and light contours correspond to 68\% and 95\%
confidence ranges, respectively.
While the time constant, $\tau$,
is treated as a free parameter ranging from 10~Myr to $\infty$,
the grid in $\tau$ contains only ten distinct values. The discrete
nature of $\tau$ translates into discrete regions of parameter
space, that would be connected smoothly if $\tau$
were allowed to vary continuously. This effect is conveyed
most clearly in the (E(B-V),Age) plots,
since both parameters systematically depend on $\tau$.
For full resolution figure, go to http://astron.berkeley.edu/$\sim$aes.}
\label{fig:mcplots}
\end{figure*}

\subsection{Systematic Uncertainties}
\label{sec:modsysunc}

Even neglecting the effects of photometric uncertainties,
parameters derived from SED fitting are subject to
significant degeneracies and systematics, primarily
because of the inability to constrain the star-formation
history \citep{shapley2001,papovich2001}. These
uncertainties apply even using the simplest
single-component models, as we do here, since the
best-fit age, extinction, and star-formation rate all
depend systematically on the $\tau$ used to describe the
star-formation history. Stellar mass estimates are more
robust in the face of uncertainties in the star-formation
history. Here we review the systematic uncertainties 
in age, extinction, star-formation rate, and stellar mass
that result from population synthesis modeling of galaxy 
SEDs at high redshift.

The inferred age, $t_{sf}$, for a given
value of $\tau$ is constrained mainly by the magnitude of
the age-sensitive Balmer break. A given Balmer break strength
indicates the relative number of A and O stars and
corresponds to older ages when larger values are assumed
for $\tau$ (i.e. a given relative number of A/O stars is
reached at later times for more gradually declining
star-formation histories).  

The inferred extinction, $E(B-V)$, also
depends systematically on the star-formation history:
declining star-formation histories with $t_{sf}/\tau>1$
with less dust extinction can produce similar $G-{\cal
R}$ and ${\cal R}-K_s$ colors to those of a model with a
larger $\tau$, smaller $t_{sf}/\tau$ and more dust
extinction. In this case, the reddening of the rest-frame
UV slope is caused by a mixture of stars with later
spectral types on average for models with smaller $\tau$,
whereas dust extinction accounts for the UV reddening in
continuous star-formation models or those with larger
values of $\tau$.  

The systematic dependence of the inferred 
SFR on the value of $\tau$ that parameterizes the star-formation
history stems from two effects. First, the star-formation
rate is derived from the extinction-corrected
UV-luminosity. It therefore depends on the best-fit value
of $E(B-V)$, which as we discussed above, depends on the
star-formation history used to model the observed SED.
Second, a conversion from extinction-corrected
UV-luminosity to SFR is required. This conversion depends
on both $t_{sf}$ and $\tau$, and for a given galaxy, is
generally an increasing function of $\tau$. The
conversion from UV-luminosity to SFR equilibrates after
about 100~Myr for models with $\tau\geq 1$~Gyr
\citep{kennicutt1998}. However, the ratio of SFR to UV
luminosity declines as a function of time for models with
smaller $\tau$. In the range of parameter space spanned
by the galaxies in our sample, $\tau=100$~Myr models
typically have UV-to-SFR conversions $1.2-1.8$ times
lower than continuous SFR models that fit the same
colors. While not likely, based on external
(spectroscopic) information, models with $\tau=10$~Myr
have much more extreme conversions from UV-luminosity to
SFR, as much as $25-1000$ times lower than the limit
approached by the models with large $\tau$. This is
easily understood since the largest $\tau=10$~Myr age
that can reasonably fit the observed colors of the
galaxies in our sample is roughly $t_{sf} \sim 100$~Myr.
At this point, the star-formation rate is roughly 20000
times lower than at $t_{sf}=0$. However, the
UV-luminosity has only declined by a factor of $\sim 20$,
because while there are no O-stars left, the lifetime of
late B-stars is roughly $100$~Myr, so they have not yet
disappeared.

Our single-component-model estimates of stellar masses
are not plagued by the same amount of systematic
uncertainty, especially when considering more realistic
exponential timescales, $\tau \geq 100$~Myr. There has
been much discussion in the literature about the value of
rest-frame near-IR photometry in measuring stellar masses
\citep{glazebrook2004,bell2003}. In fact, as shown by the
model fits to BX490 and BX505, which have very similar
rest-frame near-IR magnitudes, rest-frame near-IR
photometry alone is insufficient to constrain the stellar
mass-to-light ratio of star-forming galaxies to better than
a factor of 10 (see also \S\ref{sec:resultsmcorr}).
However, with rest-frame UV, optical, and near-IR
photometry, the near-IR $M/L$ ratio (and
therefore the stellar mass) is reasonably
well-determined, almost independent of the value of
$\tau$ used to fit the SED. With measurements of the
rest-frame UV slope (observed $G-{\cal R}$ color at high
redshift), the rest-frame UV/optical color (observed
${\cal R}-K_s$), and rest-frame UV/near-IR color
(observed ${\cal R}-\mbox{IRAC}$), the variance in
inferred near-IR $M/L$ ratio due to uncertainties in the
form of the star-formation history (for $\tau \geq
100$~Myr) is typically 10\%.  
The relationship between ($G-{\cal R}$, ${\cal
R}-K_s$, ${\cal R}-\mbox{IRAC}$) and $M/L$ ratio starts to
break down for star-formation histories with $\tau<100$~Myr. 
The systematic uncertainty in the $M/L$ ratio
therefore increases, except in the cases where $t_{sf}
\leq 10$~Myr (and $\leq \tau$ for all star-formation
histories under consideration), when all star-formation
histories yield the same best-fit stellar population
parameters. However, for most galaxies in the sample,
there is no indication that models with $\tau \leq
100$~Myr with $t_{sf}/\tau>1$ are preferred, and in some
cases such star-formation histories are actually ruled
out (see discussion in section~\ref{sec:discussmassive}).

External constraints can be used to
evaluate the likelihood that these extreme ``decaying
burst'' star-formation histories are realistic.  For
example, \citet{reddy2004} have shown that the mean
extinction-corrected star-formation rate for $z\sim2$
galaxies in the Chandra Deep Field North, inferred from
rest-frame UV luminosities and colors, assuming
$>100$~Myr of continuous star formation and the Calzetti
extinction law, agrees very well with the average
star-formation rate derived from stacked X-ray and radio
images of the same galaxies. If $\tau < 100$ had been
assumed, the average UV-derived star-formation rate would
have been several times smaller, and no longer in
agreement with the star-formation rate inferred from the
average X-ray or radio continuum flux.  In principle, the
degeneracies that affect primarily the interpretation of
the UV continuum slope can be removed if there is
external information on the number of massive stars still
present in the galaxies. For example, using observed
H$\alpha$ fluxes and/or high quality far-UV spectra, it
is relatively easy to establish whether O stars are still
present in the galaxy spectra. In general, the high
success rate in detecting H$\alpha$ emission from
UV-selected galaxies, and in particular the fact that the
star-formation rate inferred from H$\alpha$ fluxes is
never smaller than that inferred from the UV continuum
\citep[e.g.,][]{erb2003}, suggests that it must be quite
rare to catch a starburst with $t_{SF}/\tau >> 1$ and
$\tau \le 100$~Myr. Additionally, in the individual LRIS
rest-frame UV spectra with sufficient S/N, one sees
evidence for the existence of massive stars from the
presence of C~IV and Si~IV P-Cygni stellar wind features,
and, in some cases Ly$\alpha$ emission. If $\tau<100$~Myr
and $t_{sf}/\tau$ were much greater than 1 in most cases,
we would not expect to find such strong evidence for
massive O stars in the rest-frame UV spectra of these
objects.

The systematic effects discussed above do not include the
uncertainties that result from including models with
underlying short bursts of different strengths occurring
at random times in the past \citep{kauffmann2003}. We did
not add that layer of complexity to our modeling
procedure, given the difficulty we found in
distinguishing among even simple exponentially declining
and continuous star-formation histories. However, as
\citet{papovich2001} demonstrated, underlying maximally
old bursts with $t_{sf} >> \tau$ can increase the $M/L$ by
a factor of a several without having an appreciable
effect on the rest-frame UV-to-optical colors.
\citet{glazebrook2004} find that including random bursts
into the range of star-formation histories used to fit
the $VIz'K$ colors of $1.0 \leq z < 2.0$ galaxies causes
the derived masses to increase by less than a factor of
two, although the amount by which one might be
underestimating the true stellar mass will depend on the
observed galaxy colors. We return to this
question in section~\ref{sec:resultsmcorr}.

\begin{figure*}
%\plotone{seds1.eps}
\epsscale{1.2}
\plotone{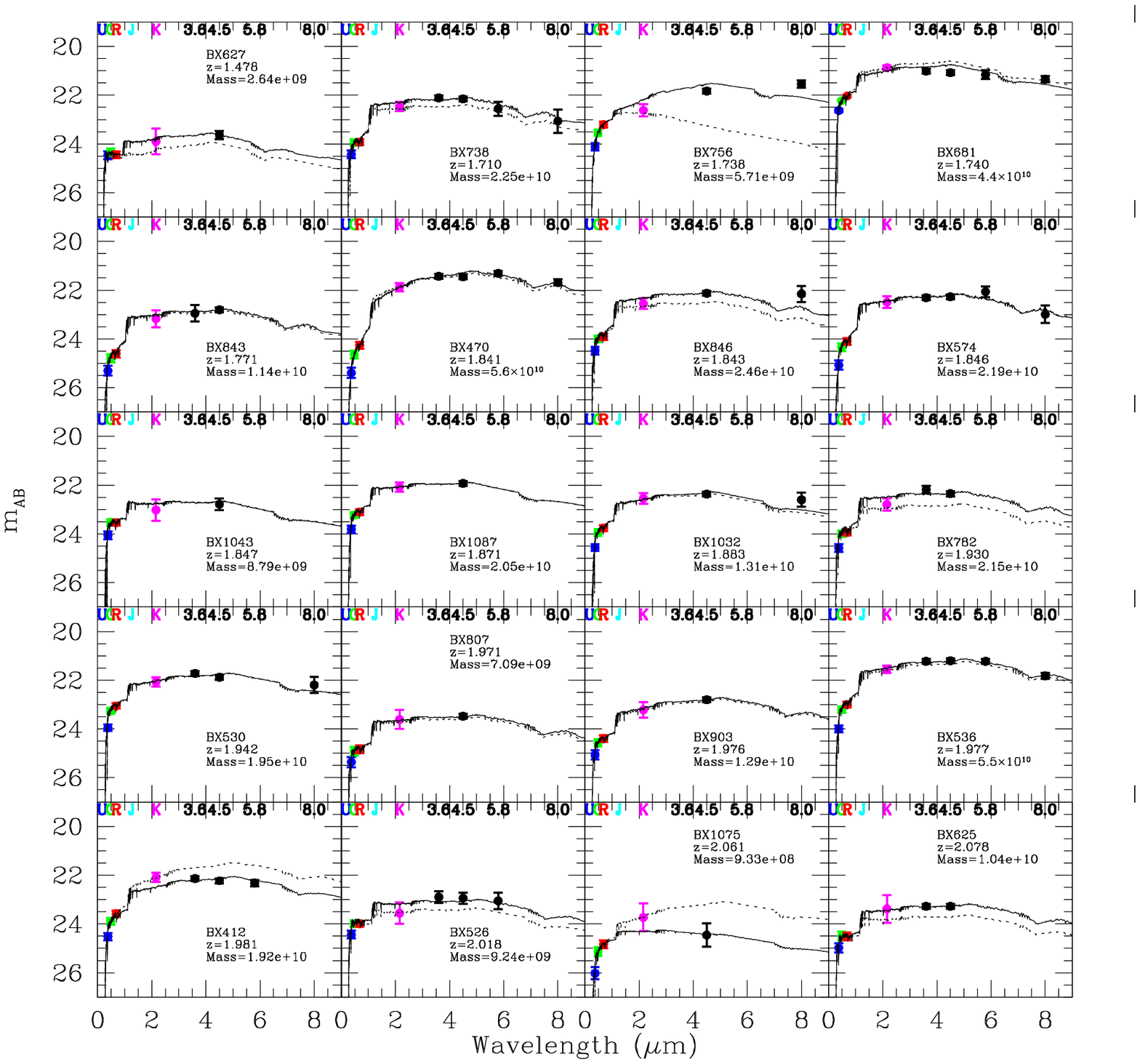}
\caption{
Observed and model SEDs for all 72 of the galaxies with
spectroscopic redshifts and measurements in the $K_s$
band and at least one IRAC channel, sorted by redshift
from lowest to highest. The numbers for the inferred
stellar mass are taken from the best-fitting
exponentially declining model, which is shown shown with
a solid curve. The dotted curve represents the best-fit
model when the IRAC data points are excluded.
}
\epsscale{1.0}
\label{fig:seds1}
\end{figure*}
\newpage

\begin{figure*}
\addtocounter{figure}{-1}
%\plotone{seds2.eps}
\epsscale{1.2}
\plotone{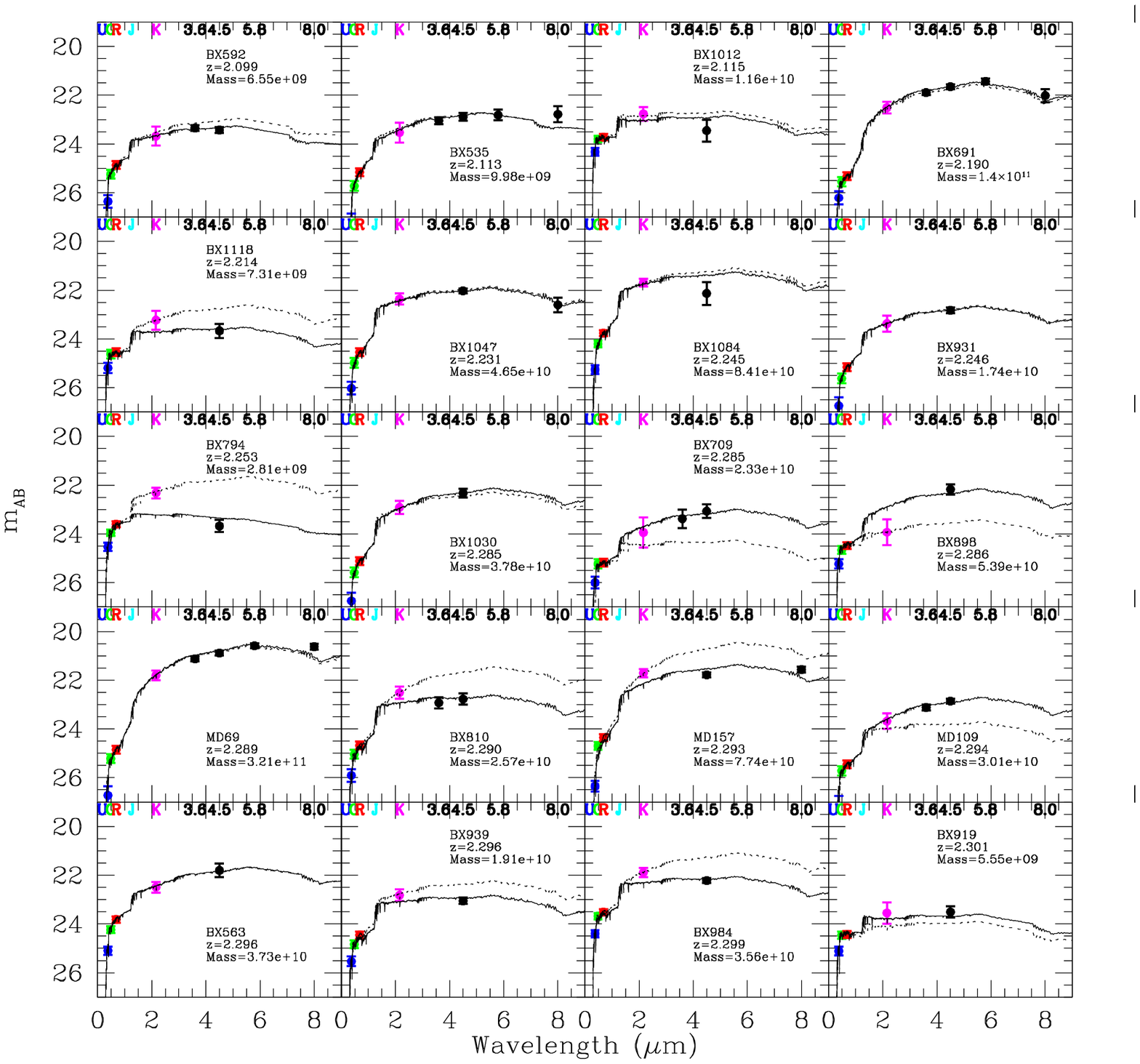}
\caption{
(continued)
}
\epsscale{1.0}
\label{fig:seds2}
\end{figure*}
\newpage

\begin{figure*}
\addtocounter{figure}{-1}
%\plotone{seds3.eps}
\epsscale{1.2}
\plotone{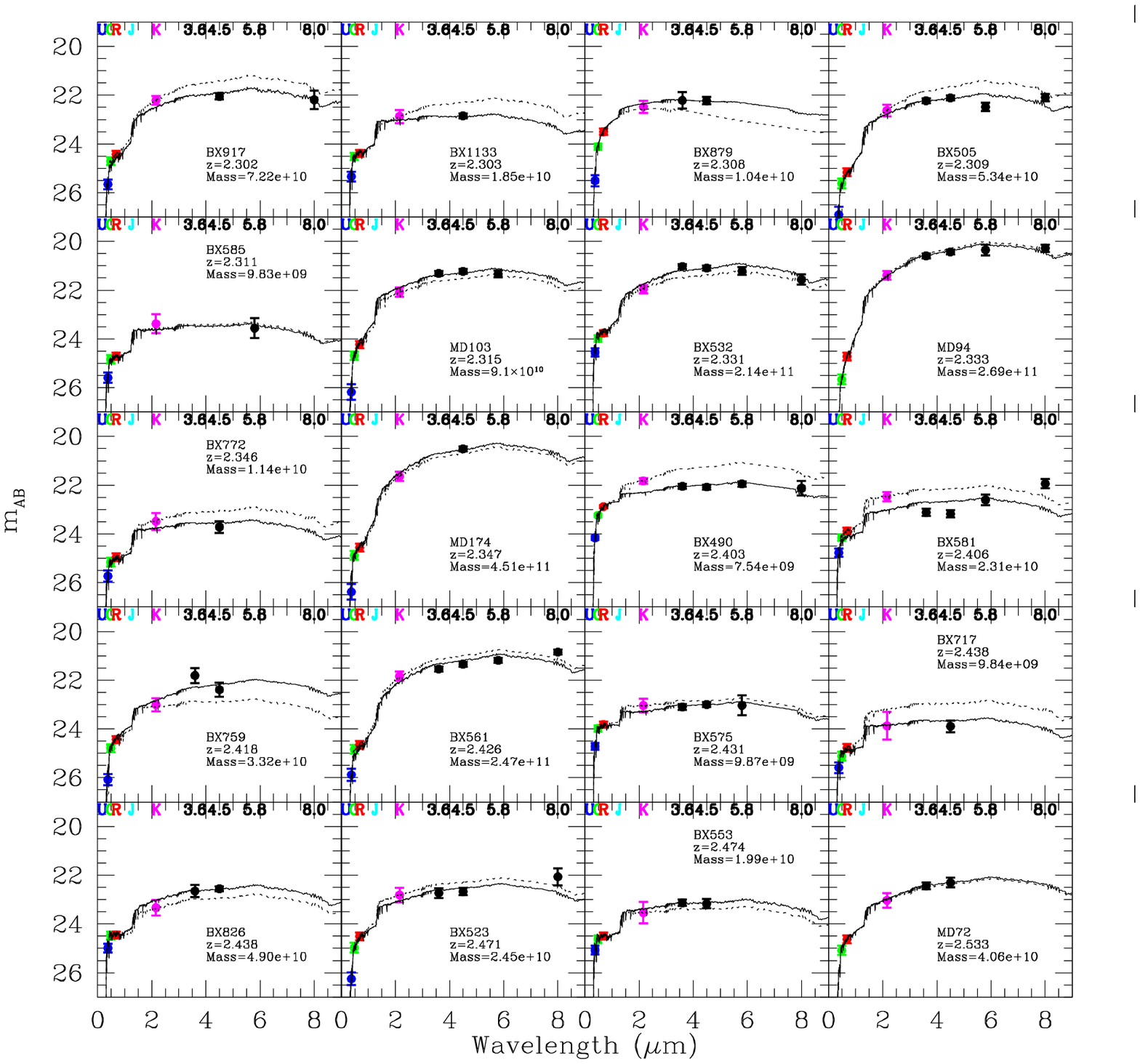}
\caption{
(continued)
}
\epsscale{1.0}
\label{fig:seds3}
\end{figure*}
\newpage

\begin{figure*}
\addtocounter{figure}{-1}
%\plotone{seds4.eps}
\epsscale{1.2}
\plotone{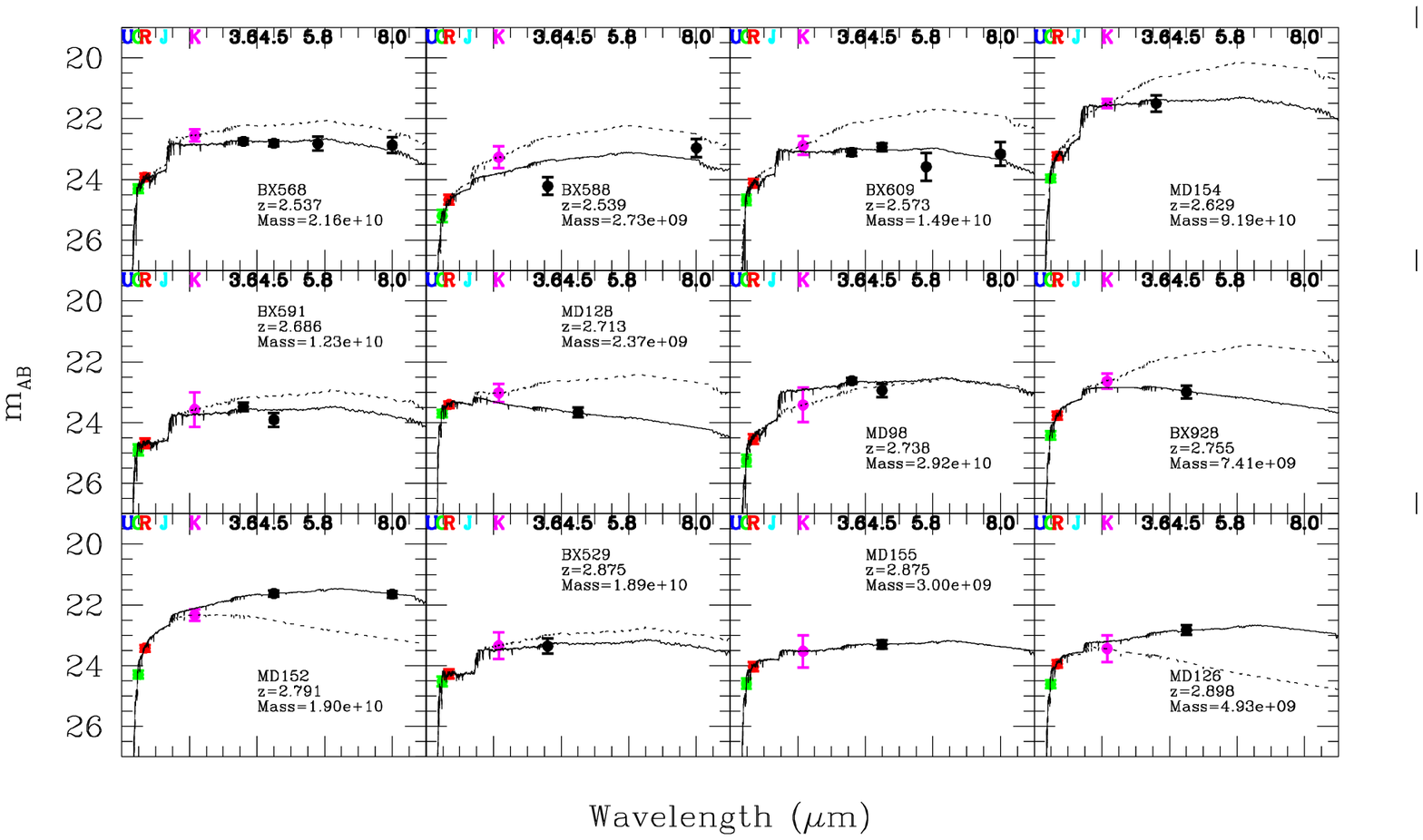}
\caption{
(continued)
}
\epsscale{1.0}
\label{fig:seds4}
\end{figure*}
\newpage

\subsection{Monte Carlo Confidence Intervals based on Photometric Errors}
\label{sec:modmcunc}

In addition to the best-fit stellar population
parameters for each galaxy, we also found the
parameter confidence intervals allowed by the photometric
uncertainties. A large set ($n=1000$) of fake
SEDs was generated for each galaxy, by perturbing the observed 
SED in a manner consistent with the estimated photometric error bars 
(sections~\ref{sec:photoptir} and \ref{sec:photirac}).
Each perturbed SED was fit in exactly the same manner as the observed
SED, and a set of best-fit stellar population parameters
was chosen to minimize $\chi^2$ with respect to
the perturbed SED. 

We determined the confidence intervals over $E(B-V)$,
$t_{sf}$, SFR, $M_{\ast}$, and $\tau$, such that 68.3\%
and 95.4\% confidence intervals contain the corresponding
fractions of the Monte Carlo realizations.
More than half of the galaxies in our sample have 95.4\%
confidence intervals that span all $\tau$ values, 
indicating the difficulty of distinguishing
among the different star-formation histories with
broad-band photometry alone. In practice, the most
useful number we derive from the Monte Carlo
confidence intervals for each galaxy is the uncertainty
on the stellar mass. For this quantity, it is important to
know not only the random effects of photometric errors,
but also the systematic effect of allowing $\tau$ to
vary. The $1\sigma$ Monte Carlo stellar mass uncertainties in
column 7 of Table 3, derived from simulations in which
$\tau$ was allowed to vary, reflect the uncertainty
in $\tau$. We also list the Monte Carlo stellar mass
uncertainties for CSF star-formation histories in Table 3.

To demonstrate the covariance of different parameters,
and at least some of the range of stellar populations
present in our sample, we show in figure~\ref{fig:mcplots}
the $t_{sf}$ vs. $E(B-V)$ and $M_{\ast}$ vs.
$t_{sf}$ confidence intervals derived from Monte
Carlo simulations. These confidence intervals do not
assume a fixed value for $\tau$, but are the results 
of simulations in which $\tau$ was allowed to vary.
The confidence intervals displayed include those of
BX879 ($z=2.308$), which appears to be a very
young ($t_{sf} \leq 10$~Myr), dusty stellar population
with high instantaneous star-formation rate and low
stellar mass, all of which 
parameters are independent of
$\tau$ because of the extremely young age;  
BX691 ($z=2.190$), an evolved stellar population with
$t_{sf}/\tau>1$, $\tau$ constrained to be between 
$0.5$  and $2$~Gyr, moderate current
star-formation rate and large stellar mass; 
and BX563 ($z=2.296$), for which there are no constraints on
the star-formation history, but the stellar mass of which
is close to the sample median. The large extent of the
BX563 confidence interval 
in $t_{sf}$ vs. $E(B-V)$ space demonstrates
some of the fundamental limitations of population
synthesis modeling.

\subsection{Caveats to the Model Results}
\label{sec:modprob}

The goal of population synthesis modeling
is to infer physical quantities from observed
galaxy magnitudes and colors, in particular
the stellar mass. In the best-case scenario we would
also like to place constraints on the timescale
of the current star-formation episode, and therefore
the level of dust extinction and the
dust-corrected star-formation rate. There
are certain limitations of the modeling procedure,
however, that are worth pointing out, in addition
to the systematic uncertainties discussed in 
section~\ref{sec:modsysunc}.

First, it is not always possible to find a 
statistically acceptable fit to the observed
galaxy colors, especially using relatively
simple models. Seven galaxies in the sample
have best-fit models with significantly higher
$\chi^2$ values than those of the rest of the sample.
Two of these galaxies (BX490
($z=2.403$) and BX794 ($z=2.253$)) have large ${\cal
R}-K_s$ residuals that are likely to be explained in
terms of contamination from $\Ha$ emission in the
$K_s$ band.  Another galaxy, BX561 ($z=2.426$), has a
good fit except for the IRAC $8.0$~$\mu$m point, which
yields a significant positive residual, possibly due to
the presence of an obscured AGN (see
\S\ref{sec:resultsagn}).  MD157 shows rest-frame UV
spectroscopic evidence for hosting a broad-line AGN,
which probably contaminates the $K_s$ flux (strong $\Ha$)
and IRAC $8.0$~$\mu$m band. The SEDs of the remaining three
galaxies (BX581 ($z=2.406$), BX756 ($z=1.738$), and BX681 ($z=1.740$))
are very difficult to understand, 
with both large $K_s$ and IRAC residuals that
are both positive and negative, and don't stem
from obvious irregularities in any of the
imaging data.

A second issue is the apparent discrepancy
between model parameters based on the best fit to
broad-band photometry and those inferred from
rest-frame UV spectra, which applies for
a small number of galaxies in the sample:
BX1032 ($z=1.883$), BX1087 $(z=1.871)$, and BX536
($z=1.977$). Not only did we obtain statistically
acceptable fits to the observed photometry of these 
galaxies, but also
the results of the Monte Carlo simulations appear to rule out
star-formation histories with $\tau > 200$~Myr, and
to favor best-fit $t_{sf}/\tau$ significantly greater than
unity, with quite modest star-formation rates of a few
$M_{\odot} \mbox{yr}^{-1}$. The broad-band photometry of
these galaxies implies star-formation histories
in which the bulk of stellar mass formed in the past, and
the current star-formation rate is much lower than the
past average. We also have rest-frame UV Keck~I/LRIS
spectra for these galaxies, which can be used as a
consistency check on the stellar population parameters
derived from the SED-fitting procedure. The spectra of
BX1032 and BX1087 both show prominent C~IV P-Cygni
features, which are produced in the winds of massive
stars, associated with active ongoing star formation. The
rest-frame UV spectra of BX1087 and BX536 both contain
broad He~II emission, which is produced by Wolf-Rayet
stars, another indication of a young stellar population.
Finally, the spectrum of BX1032 exhibits nebular emission
lines such as CIII] and He~II, further indications of
active current star formation. In short, the rest-frame
UV spectra of these $t_{sf}/ \tau >> 1$ galaxies contain
features that are not consistent with the best-fit models
used to describe the rest-frame UV to near-IR SED, which
is all the more striking since the models appear fairly
well-constrained. 

Clearly, there are cases in which the
simple, single-component SED-fitting procedure fails 
to capture the true nature of
the galaxy stellar population.  Rest-frame UV and optical
spectra, and X-ray, submillimeter, and radio fluxes,
provide valuable complementary information that must be
included to better-constrain all of the stellar
population parameters.  The limitations inherent in SED
fitting, without external constraints (and especially
without spectroscopic identifications), should be borne
in mind when considering the resulting stellar population
parameters. On the other hand, the stellar masses
inferred for these three galaxies are largely independent
of the star-formation history assumed, and therefore we
include them in all further discussion of stellar masses.

\section{Results}
\label{sec:results}

\subsection{The Stellar populations of $z\sim 2$ Galaxies in the HS1700 Field}
\label{sec:resultspop}

In Figure~\ref{fig:seds1}, we show the observed optical,
near-IR, and {\it Spitzer} photometry and best-fit models
for the entire sample of UV-selected $z\sim 2$ galaxies
in the HS1700 field with spectroscopic redshifts, $K_s$,
and {\it Spitzer}/IRAC photometry. The solid lines
indicate best-fit exponential \citet{bc2003} models
including the IRAC measurements, whereas the dotted lines
indicate fits to only $U_nGRK_s$ photometry. The stellar
mass inferred from the best-fit to the entire SED (including
IRAC photometry) is also shown, along with the redshift
of each object. In Table~\ref{tab:mod}, we list the
best-fit stellar population parameters for each object,
including the entire optical through IRAC SED. Parameters
are listed for the best-fit model assuming $\tau=\infty$
(CSF), and also for the exponentially declining model
that formally yielded the lowest value of $\chi^2$ when
$\tau$ was allowed to vary as a free parameter.
There is a subset of galaxies with extremely young
best-fit ages ($t_{sf} \leq 10$~Myr), for which all
values of $\tau$ yield nearly identical stellar
population parameters. These galaxies include: BX928,
BX1075, BX794, MD128, BX879, BX490, MD155, MD152, BX588,
and MD126. In such cases where $t_{sf}$ is smaller than
any of the $\tau$ values considered, the star-formation
history (as parameterized by $\tau$) becomes irrelevant.

From both Figure~\ref{fig:seds1} and Table~\ref{tab:mod},
it is clear that a large range of rest-frame UV to
rest-frame near-IR SEDs is included in our sample. In
particular, we highlight the variety of rest-frame
UV/near-IR colors, which translates directly into a large
range of inferred near-IR mass-to-light ratios.

\subsection{Distribution of Stellar Masses}
\label{sec:resultsmass}

\begin{figure}
%\plotone{masshist_dash.eps}
\plotone{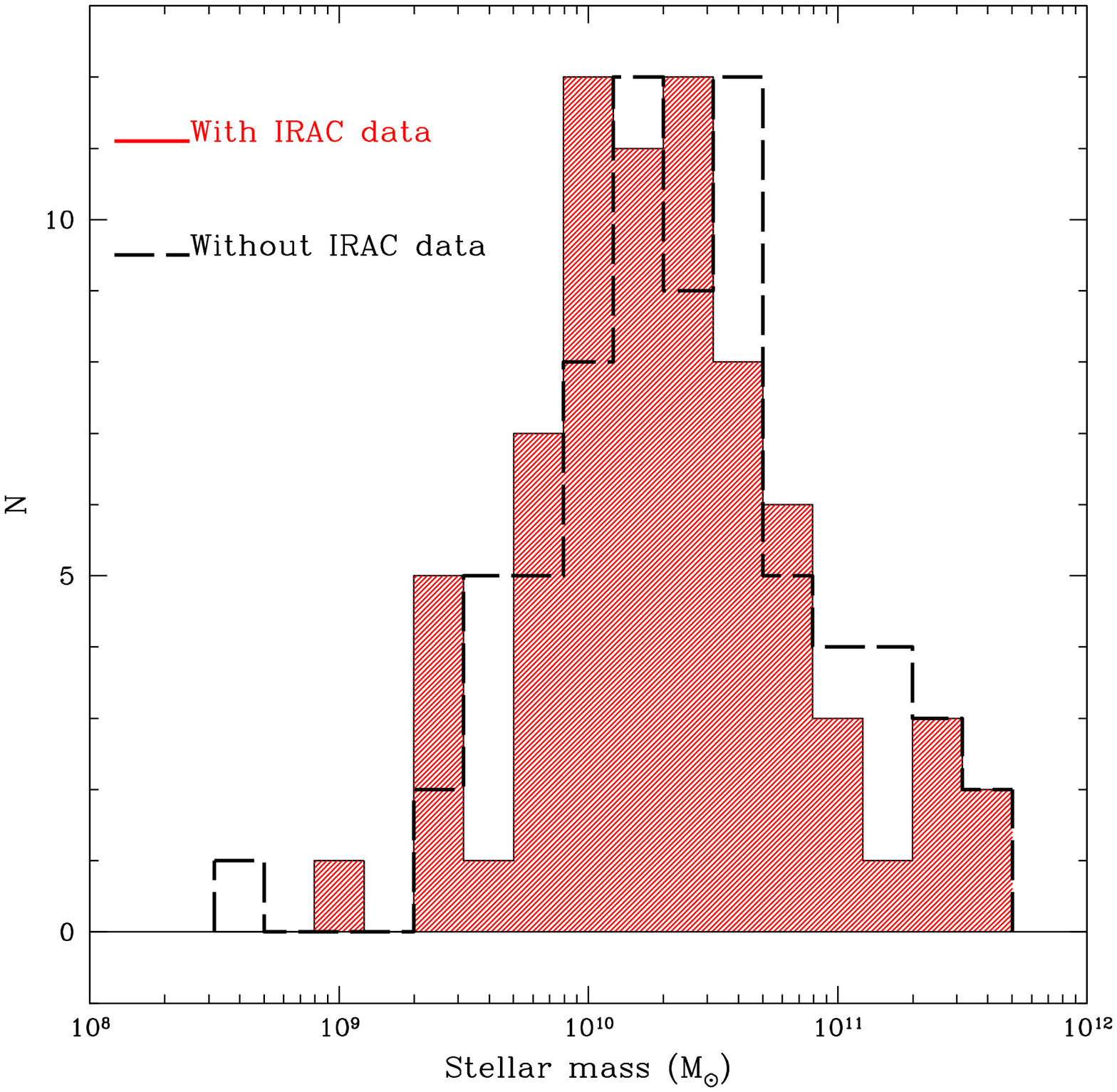}
\caption{
Histograms of best-fit stellar masses, with (shaded) and
without (unshaded, dashed) IRAC photometry. As discussed
in the text, the overall mass distribution and the total
inferred stellar mass are similar to within $\sim 15$\%,
although the uncertainties in the measurement of
individual stellar masses are smaller by a factor of
$\sim 1.5-2$ when the IRAC data are included.
}
\label{fig:mhist}
\end{figure}

\begin{figure}
%\plotone{m_vs_m.eps}
\plotone{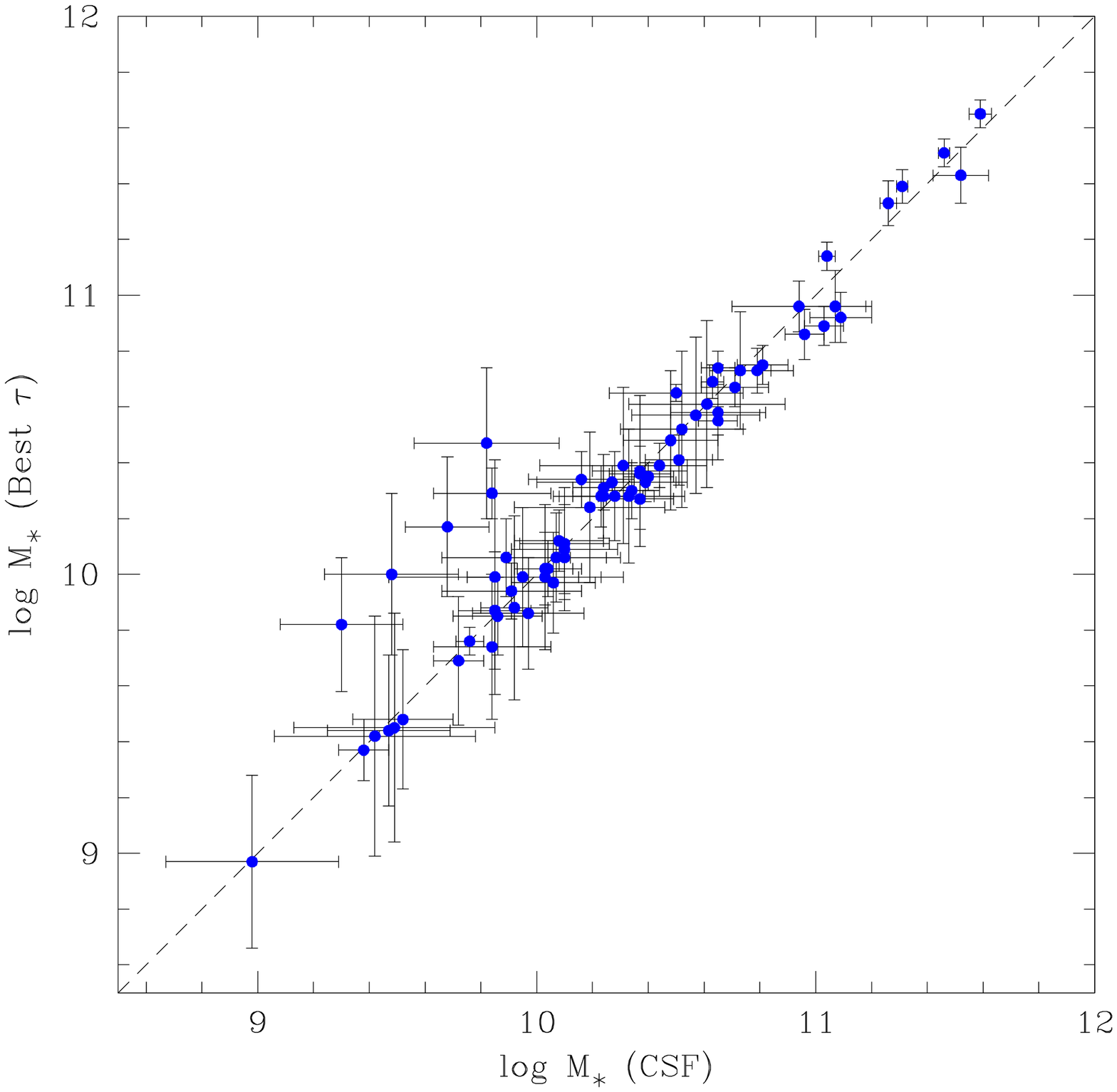}
\caption{
A comparison of the inferred stellar masses for constant
star formation (CSF) models and for the best-fit
declining star formation models (best $\tau$) for the 72
galaxies with IRAC measurements.  The error bars are as
presented in Table~\ref{tab:mod}, and are based on Monte
Carlo modeling given the photometric errors. Note that
there are no obvious systematic differences between the
CSF and $\tau$ models, although in some cases the $\tau$
models result in significantly better agreement with the
observed data.
}
\label{fig:mcsfmtau}
\end{figure}

Figure~\ref{fig:mhist} shows the distribution of stellar
masses inferred from the optical--IRAC SEDs. The mass for
each object is taken from column 7 of
Table~\ref{tab:mod}. The nearly log-normal distribution
of stellar mass can be characterized by a mean and
standard deviation ${\rm log} M_{\ast} = 10.32\pm0.51$
M$_{\odot}$. For the CSF models (the results of which are
also compiled in table~\ref{tab:mod}), we find a
distribution ${\rm log} M_{\ast}(CSF)=10.30\pm0.53$.  
Figure~\ref{fig:mcsfmtau} compares the stellar mass
obtained assuming the best-fit $\tau$ model to the
stellar mass implied by the best-fit CSF models. The
agreement, both in terms of the mass distribution and the
masses of individual objects, is excellent, with no
obvious systematic trend or offset.

For some purposes, it is of interest to break down the
inferred stellar mass distribution as a function of
observed $K_s$-band magnitude, since a number of other
surveys of galaxies at similar redshifts are selected in
that band. The results are summarized in
Table~\ref{tab:MvsK}, where the $K_s$ limits are chosen
to correspond to the limits imposed by some of these
other surveys. As expected, there is a clear trend of
increasing inferred stellar mass with increasing
rest-frame optical luminosity, although, as discussed
below, there is significant scatter in rest-optical $M/L$
at a fixed mass or luminosity.

The typical (median) Monte Carlo uncertainties in
individual stellar mass estimates are $\sim 30-35$\%,
derived from simulations in which $\tau$ was allowed
to vary as a free parameter.
If we model all galaxies in the sample with
CSF star-formation histories, we find that the median
Monte Carlo stellar mass uncertainty is $\sim 35$\%.
These values indicate that,
for the same photometry and uncertainties, CSF models
actually yield stellar mass uncertainties at least as large as the
ones derived when $\tau$ is allowed to vary as a free
parameter.

\subsection{Results with and without {\it Spitzer}/IRAC Photometry}
\label{sec:resultsiracnoirac}

One of the fundamental limitations of the analysis in
\citet{shapley2001} and \citet{papovich2001} was the the
inability to discriminate among different star-formation
histories with only rest-frame UV and optical photometry.  
We have shown that, even with the addition of {\it
Spitzer}/IRAC photometry, this inability to constrain
star-formation histories persists for most galaxies in
our $z\sim 2$ sample. The broad-band SEDs are
particularly degenerate with respect to age, dust
content, and star-formation rates, while the inferred
stellar mass suffers from much less systematic
uncertainty.  Another question raised in the pre-{\it
Spitzer} era was whether or not rest-frame near-IR
photometry of $z \geq 2$ galaxies would uncover
considerably more stellar mass from old bursts than had
been inferred from SEDs whose longest wavelength
measurement was the ground-based $K_s$ band,
corresponding to the rest-frame optical at the redshifts
of interest here.  Such stellar composite populations
might have been revealed through IRAC fluxes in excess of
the rest-frame near-IR extension of model fits to the
rest-frame UV through optical SEDs.

\begin{figure}
%\plotone{irac_vs_noirac.eps}
\plotone{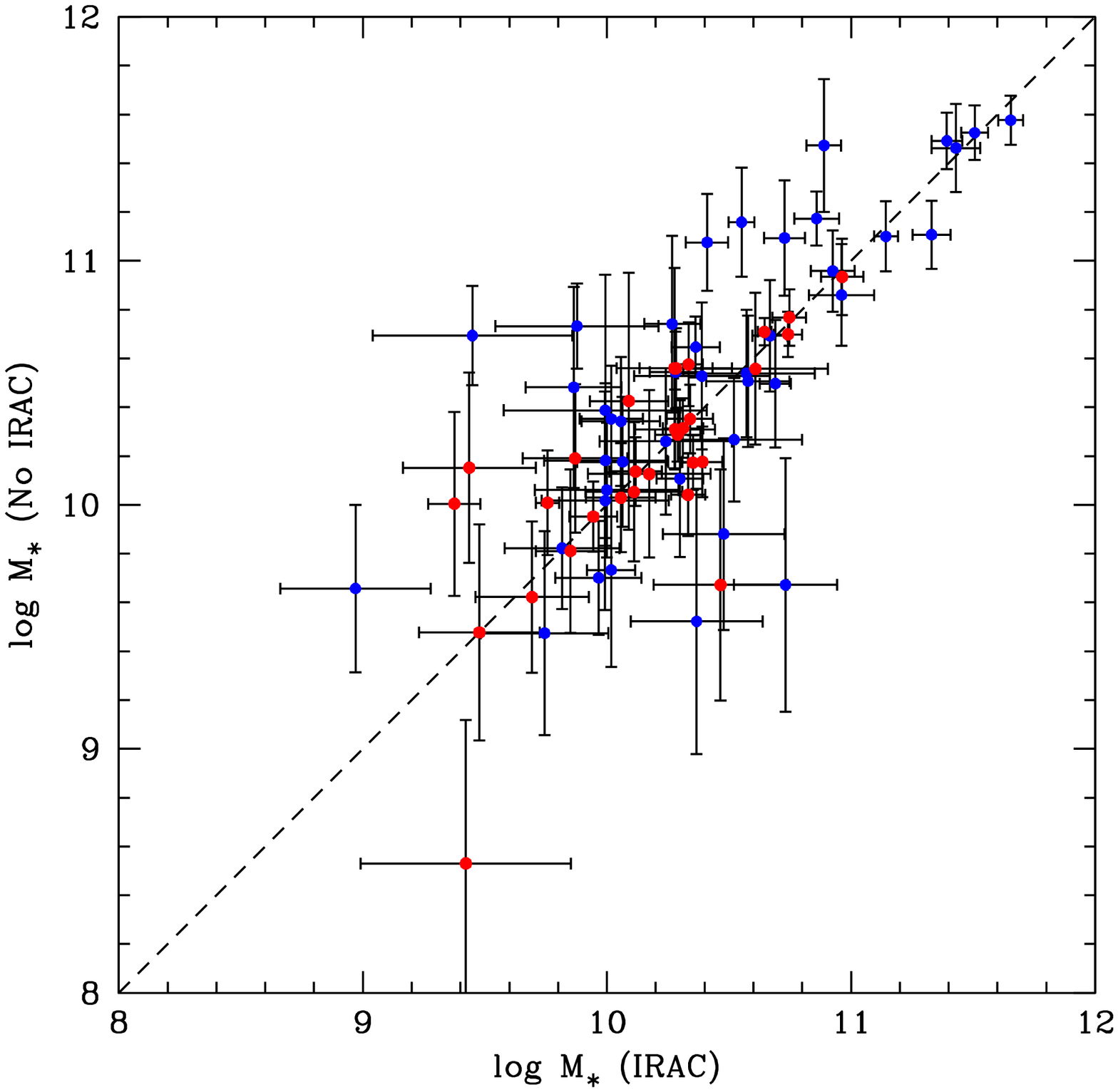}
\caption{
A comparison of the inferred stellar masses for the
best-fit declining star formation models (best $\tau$)
for the 72 galaxies with IRAC measurements, derived with
and without IRAC photometry. The error bars are based on
Monte Carlo modeling given the photometric errors. There
are very few galaxies for which the stellar mass is
significantly underpredicted when IRAC photometry is
excluded. It is more common for the no-IRAC fits to
biased towards higher stellar masses, in at least some
cases because of the extra contribution to the $K_s$-band
light by $\Ha+\mbox{[NII]}$ emission. The red points are
galaxies whose redshifts place $\Ha$ outside of the $K_s$
band, while the blue points are objects with $2.016 < z <
2.504$ for which significant contamination of the $K_s$
measurement by $\Ha$ emission is possible.
}
\label{fig:iracvnoirac}
\end{figure}

Figure~\ref{fig:iracvnoirac} shows the relationship
between $M_*(\mbox{No IRAC})$, the stellar mass inferred
when the IRAC photometry is excluded from the fit, and
$M_*(\mbox{IRAC})$, the stellar mass derived using IRAC
photometry. Figure~\ref{fig:mhist} displays this
relationship in one dimension, with the distribution of
$M_*(\mbox{No IRAC})$ indicated as an unshaded histogram
in front of the shaded histogram of $M_*(\mbox{IRAC})$.
What both figures demonstrate is that, on average, the
stellar mass inferred with and without IRAC data agree
without a large systematic offset, though the scatter in
this relationship is significant and larger than the
typical (formal) uncertainties in the individual stellar
mass estimates using the full SED.  Table \ref{tab:MvsK}
compares these inferred stellar mass distributions as a
function of limiting observed $K_s$ flux.

The average of log $M_*(\mbox{No IRAC})$ - log
$M_*(\mbox{IRAC})$ =0.06, with a standard deviation of
0.35 dex; the median difference is only 0.01 dex.  We
find only five galaxies whose optical--IRAC SEDs imply
stellar masses more than a factor of two larger than that
inferred from the optical--$K_s$ SEDs. The most dramatic
such example is BX898, whose observed $4.5 \mu$m flux is
$\sim 3.5$ times larger than that predicted by the fit to
the optical--$K_s$ SED alone. The inferred stellar mass
is $\sim 12$ times larger if the IRAC $4.5 \mu$m point is
included in the SED fit. Unfortunately, BX898 has a
significant IRAC measurement only at $4.5 \mu$m, and
contamination of this measurement (or, a statistical
outlier in the observed $K_s$ measurement) cannot be ruled
out.

It is more common for the fit to the optical--$K_s$ SED
to {\it overpredict} the stellar mass relative to that
inferred from the optical--IRAC SED. We believe that this
discrepancy arises in most cases because of the
contamination of the $K_s$ continuum flux by
$\Ha+\mbox{[NII]}$ line emission at $2.0 \leq z \leq
2.5$. The addition of the IRAC photometry to the SEDs of
galaxies with strong $\Ha$ emission sometimes reveals the
$K_s$ data point as a positive outlier relative to the
overall SED fit (e.g. BX794 and BX490), and the best-fit
Balmer break is smaller than the observed ${\cal R} -
K_s$ color alone would imply. In the absence of IRAC
photometry, the $K_s$ measurement strongly influences the
inferred stellar mass because it is the only proxy for
the strength of the Balmer break, resulting in a bias
toward somewhat higher values.  
In support of this interpretation, for the four galaxies in the
HS1700 $K_s$/IRAC pointings with published $\Ha$
measurements, we find an average ratio of 1.55 between
$M_*(\mbox{No IRAC})$ and $M_*(\mbox{IRAC})$ when no
correction for $\Ha$ emission is applied. When $\Ha$
emission is corrected for, we find an average
$M_*(\mbox{No IRAC})/M_*(\mbox{IRAC})$ of 1.11, i.e.
closer to unity.  More complete $\Ha$ surveys are
necessary to remove the $K_s$ emission line bias,
especially in order to model objects for which $K_s$ is
the reddest available band \citep{erb2005}.

Without IRAC photometry, we find typical Monte Carlo
stellar mass uncertainties of $50-60$\% for CSF models,
which is $\sim 1.5$ times larger than the typical
stellar mass uncertainties estimated when including the
IRAC photometry. When $\tau$ is allowed to vary, the
typical stellar mass uncertainty is $60-70$\%, $\sim 2$
times larger than the typical uncertainty inferred with
IRAC photometry. Thus, while the lack of IRAC photometry
does not appear to cause a large bias in the stellar mass
estimates, the inclusion of IRAC photometry significantly
reduces the formal uncertainties.

While we have noted the possibility of H$\alpha$
contamination of the $K_s$-band measurements for the
$\sim$60\% of the current sample having redshifts where
this is a potential problem, the {\it total} inferred
stellar mass for the full sample of 72 galaxies differs
by only $\sim 15$\% when IRAC measurements are included
or excluded (in the sense that the inferred total mass is
higher without the IRAC data included). Thus, while the
addition of longer-wavelength data leads to significantly
smaller uncertainties in the estimates of the stellar
masses of individual galaxies, the IRAC data do not
appear to be essential for obtaining reasonable stellar
mass estimates for galaxies at $z \sim 2$, particularly
when statistics of large samples are most relevant.

\subsection{Mass-to-Light Ratios versus Wavelength}
\label{sec:resultsmcorr}

\begin{figure}
%\plotone{K_vs_m.eps}
\plotone{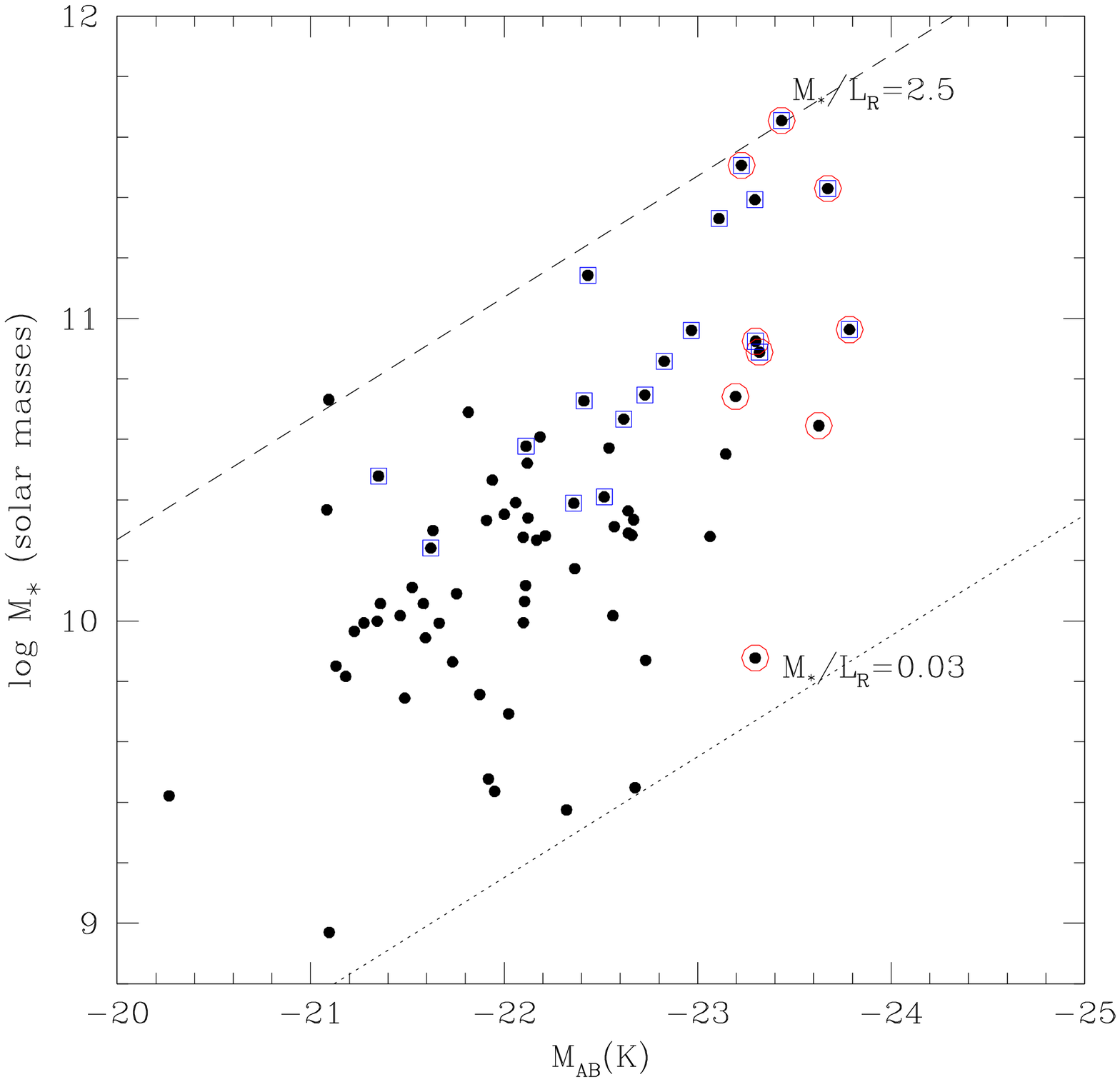}
\caption{Inferred stellar mass for the best-fit model,
versus the absolute magnitude at $K_s$ ($2.155 \mu$m)
Note that the scatter in stellar mass/light ratios at a
given $K_s$ luminosity (rest-frame R band at the mean
redshift of the sample) can be as large as a factor of
70.  The 9 objects in the sample with $K_s \le 20$ (Vega
normalized) are indicated with open circles surrounding
the points, while the reddest quartile of the sample with ${\cal R}-K_s>3.5$ are
indicated with open squares.  Dashed and dotted lines bracket the range of stell
ar $M/L$
in the sample, and are given in solar units, evaluated
at rest-frame $R$ band.  The very largest optical stellar $M/L$ seen
in $z \sim 2$ galaxies approaches that of galaxies in the
present-day universe, while the average value is $\sim 5$
times smaller. }
\label{fig:Kvm}
\end{figure}

\begin{figure}
%\plotone{ch2_vs_m.eps}
\plotone{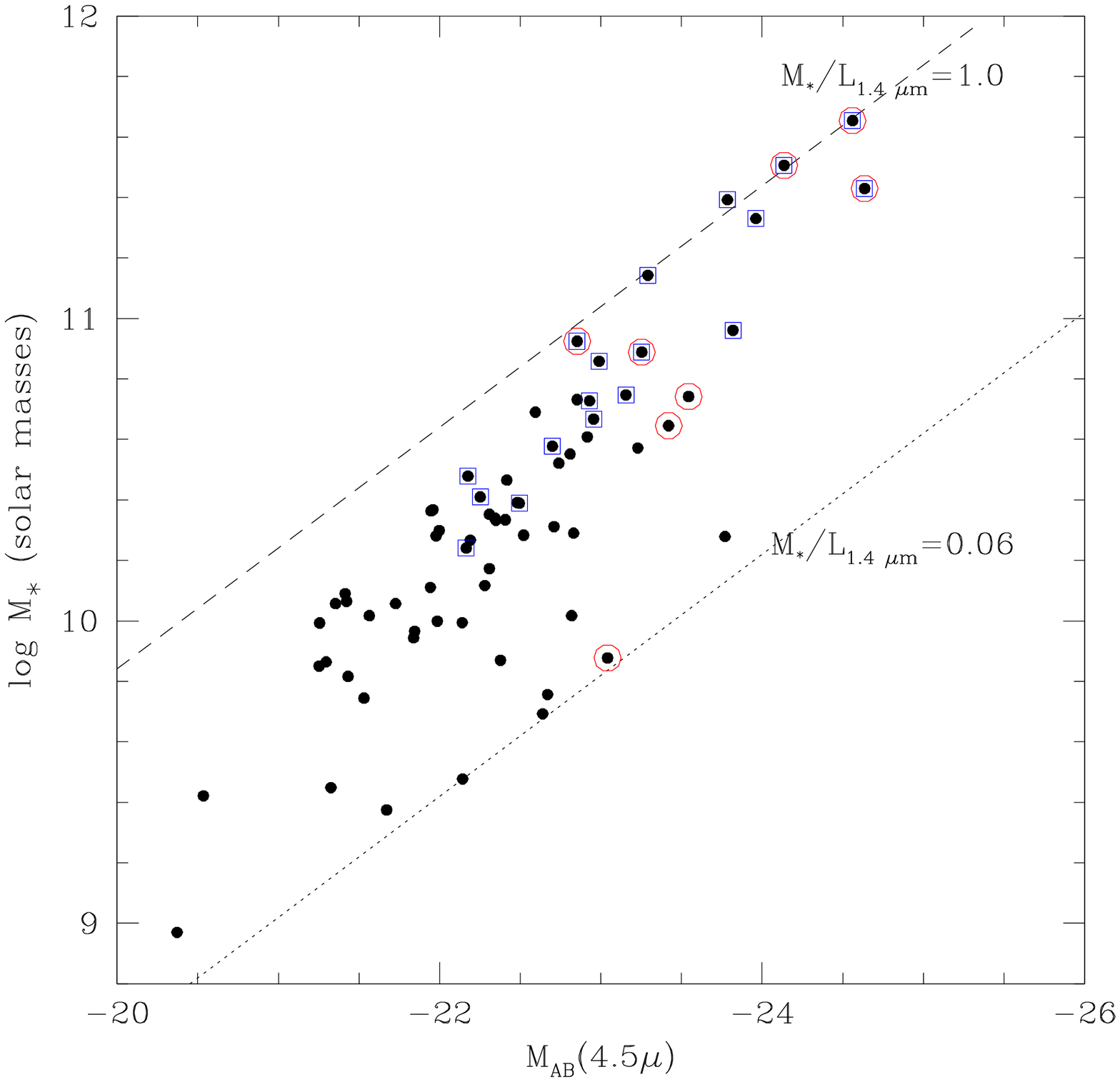}
\caption{Inferred stellar mass for the best-fit model,
versus the absolute magnitude at 4.5$\mu$m, equivalent to
the rest-frame 1.4$\mu$m luminosity at the mean redshift
of the sample. As for figure~\ref{fig:Kvm}, the objects
with $K_s \le 20.0$ are circled and those with ${\cal R}-K_s > 3.5$
are indicated with open squares.
Dashed and dotted lines bracket the range of stellar $M/L$
in the sample, and are given in solar units, evaluated
at rest-frame $1.4\mu$m. While the
scatter in stellar $M/L$ is very much reduced compared to
that measure in the $K_s$ band, it can still be as large as a factor of 10
at a given 4.5$\mu$m luminosity.
The $z \sim 2$ sample spans a range of a factor of 15, with the
most massive galaxies having attained a near-IR $M/L$
comparable to present-day galaxies, but with typical
galaxies having $(M/L)_{1.4\mu m} \simeq 0.4$, a factor of $\sim 2.5$ smaller.
There is much less scatter in near-IR $M/L$ for objects with
${\cal R}-K_s > 3.5$.}
\label{fig:ch2vm}
\end{figure}

It is often asserted that estimates of stellar mass from
photometric measurements become increasingly reliable as
one obtains longer wavelength data. We have certainly
found this to be the case, as discussed above; however,
it is not true that long-wavelength observations can be
used to measure stellar masses without understanding the
recent star-formation history of the galaxy, even when
observations extend into the rest-frame near-IR as for
the IRAC observations of $z \sim 2$ galaxies. This point
is illustrated in figures~\ref{fig:Kvm} and 
\ref{fig:ch2vm}; we have chosen the observed $K_s$ band
and the observed IRAC channel 2 (4.5$\mu$m) band since
the former is the longest wavelength measurement possible
from the ground, and the latter lies close to $\sim
1.5\mu$m in the rest frame at the mean redshift of the
current sample (68/72 of the galaxies in the sample have
been measured in IRAC channel 2), where the peak in the
SED produced by older stellar populations is expected.

Figure \ref{fig:Kvm} shows the absolute magnitude of the
galaxies based on their observed $K_s$-band
magnitudes, which at the median redshift of $\langle z
\rangle = 2.29$ corresponds to 
rest-frame 0.65$\mu$m (i.e., R band). Note that while
there is clearly a correlation between absolute R-band
magnitude and stellar mass, there is a large enough range
in star-formation history among the sample that a
measurement of a particular rest-frame luminosity would
map quite poorly onto stellar mass. The dotted and dashed
curves show that there is a range in inferred $M/L$ of a
factor of $\sim 70$ among the sample. This large factor
applies even for objects that are bright in the $K_s$
band ($K_s \le 20.0$), indicated with open circles. The
very largest stellar $M/L$ values approach that of
present-day galaxies \citep{bell2003}, but the
average value is $\sim 5$ times smaller, $\langle (M/L)_R
\rangle = 0.53$ (where both $M$ and $L$ are in solar
units, and $L$ is evaluated in the rest-frame $R$ band).  
The situation is improved at $4.5\mu$m
($\sim$ rest-frame 1.4$\mu$m), as shown in
figure~\ref{fig:ch2vm}; the scatter in the stellar $M/L$
is a factor of $\sim 15$, with indications that the
scatter decreases for the most luminous galaxies in the
rest-frame near-IR, which tend to have $M/L$ very close to
that of present-day galaxies. The scatter in $M/L$ is
also greatly reduced when the sample is limited to objects
with ${\cal R}-K_s > 3.5$ (the reddest quartile), indicated
with open squares. Still, among the total sample
studied here, the average value is $(M/L)_{1.4\mu{\rm m}}
=0.38$, roughly 2.5 times smaller than typical in the
local universe at the same rest wavelengths. The local
$M/L$ ratios were derived with an IMF containing fewer
low-mass stars than are found in the standard $0.1-100 M_{\odot}$
Salpeter IMF, used for the current work. Adopting
the ``diet'' Salpeter IMF used in \citet{bell2003} would lower the
average $z\sim 2$ $M/L$ ratios by a factor of $\sim 1.5$.

\begin{figure}
%\plotone{rmk_vs_m.eps}
\plotone{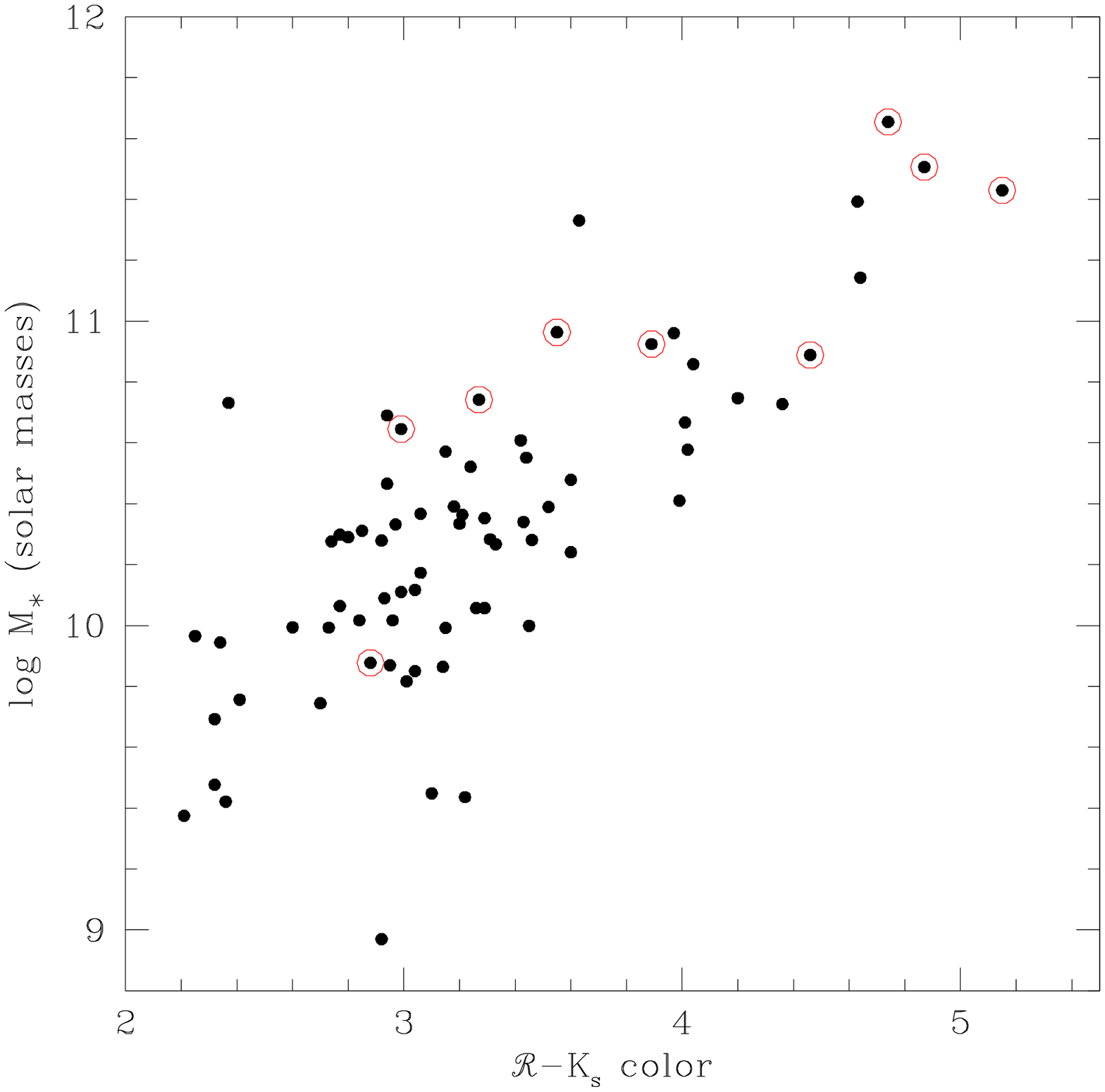}
\caption{The relationship between inferred stellar mass
and the observed ${\cal R-K}$ color for the sample of 72
galaxies with $\langle z \rangle = 2.26\pm 0.30$. There
is a very strong correlation between inferred $M_{\ast}$
and ${\cal R}-K_s$ color, particularly for ${\cal R}-K_s
> 3.5$. Again, objects with $K_s \le 20.0$ are circled.}
\label{fig:rmkvm}
\end{figure}

A simple interpretation of these results is that gauging
the importance of current or recent star formation on the
galaxy luminosities, even in the rest-frame near-IR,
depends on the observations of the full UV-to-near-IR SED
of the galaxy. That is, one cannot go directly from
near-IR luminosity to stellar mass without knowledge of
the likely contribution of young stars to that
luminosity, which can only come from measurements at
shorter wavelengths.  The region of the spectrum that is
most sensitive to the ratio between the current star
formation and the integral of past star formation is
between the UV and the visual, conveniently measured by
the observed ${\cal R}-K_s$ color in the current sample.
This point is illustrated in figure \ref{fig:rmkvm},
which shows that the inferred stellar mass is remarkably
well-correlated with ${\cal R}-K_s$. In fact, the
correlation between ${\cal R}-K_s$ color and inferred
stellar mass is almost as significant as the
correlation between stellar mass and
$M_{1.4\mu{\rm m}}$-- using a Spearman test, the
significance of the correlations are 5.9$\sigma$ and
6.4$\sigma$, respectively -- and more significant 
than the correlation between stellar mass and rest-frame
$0.65 \mu$m absolute magnitude $(5\sigma)$.
One implication of this result is
that, in the absence of IRAC data,
it may be more effective to use 
${\cal R}-K_s$ color rather then $K_s$ flux
to select $z \sim 2$ galaxies with large stellar masses.
Another implication is that there are few galaxies that
have large inferred stellar masses at $z \sim 2$ without
showing evidence for long star-formation histories 
\citep[cf.,][]{shapley2004} that are suggested by red rest-frame
2000 -- 6500 \AA\ colors.

Of course, caution is required in assessing the true
significance of correlations that are based on the model
results and not on directly measured quantities; such is
the case for all stellar mass estimates. To some extent
the results in this section simply reflect how the
observed SED is used to infer stellar mass by the
modeling software, as a single smoothly declining or
continuous episode of star formation that dominates
the emission at all wavelengths from rest-frame UV through near-IR.
In such a framework, the ratio of rest-frame UV to 
optical or near-IR light indicates the stellar $M/L$ 
in a fairly robust manner (see section~\ref{sec:modsysunc}).
We can obtain a different result for the stellar mass 
if we instead understand the galaxy UV-to-near-IR SED
as the superposition of a young, roughly continuous episode of
star formation and an old burst that peaked sometime in the past.
In the most extreme case of a 
maximally-old burst and a very young ($t_{sf} \leq 10$~Myr) continuous
episode, the old burst will dominate the rest-frame near-IR light,
the young episode will dominate the emission
at rest-frame UV wavelengths, and the maximum possible stellar mass
will be inferred from the rest-frame near-IR luminosities. 
While this scenario is rather unrealistic, it does provide an 
extreme upper limit on the stellar mass implied by the
rest-frame near-IR light. 
Less extreme multiple-component models
provide a natural representation of the episodic star-formation histories
that arise in a hierarchical model of galaxy formation, and will
always result in {\it larger} stellar mass estimates than
those derived from the single-component modeling presented 
thus far in this work. 

In order to set a rough upper limit on the stellar masses
implied by the broad-band SEDs, we used a technique very
similar to that of \citet{daddi2004}, which is designed
to infer the maximum $M/L$ ratio from the rest-frame
near-IR data points. This method accounts for most of the
light at those wavelengths with a maximally-old burst, on
top of which is superimposed a young stellar population
that accounts for most of the rest-frame UV emission. The
ages of the maximally old models range from 2.2-4.3~Gyr
for the galaxies in our sample. We first fit a young
($t_{sf} \leq 10$~Myr) CSF model to the observed
photometry, normalizing this model to the ${\cal R}$
magnitude. We then subtracted this young model from the
observed photometry, normalized the maximally old
$\tau=100$~Myr burst to the residual IRAC magnitudes,  
and summed the stellar mass from the two components.
This sum is dominated by the mass of the old component,
and is typically $\sim3$ times larger than the single-component mass. 
The correlation between $M_{1.4\mu{\rm m}}$ and stellar mass remains significant
at the same level as for the single-component models.
While ${\cal R}-K_s$ color is still significantly
correlated with stellar mass (at the $\sim 4~\sigma$ level),
the correlation is less significant than for the
case of the single-component models discussed above. Specifically,
there is much more scatter among objects with blue ${\cal R}-K_s$ colors, for which
the $M/L$ can change the most using this type of modeling. For objects with
${\cal R}-K_s > 3.5$, the inferred $M/L$ and the correlation between
stellar mass and ${\cal R}-K_s$ color are fairly insensitive to the
type of model used. 

It is easy to see that the amount of additional
stellar mass that could be ``hidden'' by the current star
formation episode is strongly dependent on the
observed ratio of UV to optical/near-IR light. For the
bluest objects, the ratio of true stellar mass to
inferred stellar mass could be  $\geq 5$
while redder galaxies (except
in extreme cases of dust obscuration) could not hide much
additional stellar mass.  For the galaxies in the sample
with the largest stellar masses derived from
single-component fits, the two component modeling
provides stellar mass upper limits only $10-40$\% higher
than the original estimates.
Thus, in general, objects which already have large
inferred stellar masses and old inferred ages cannot be
much altered by more extreme assumptions, while objects
that are completely dominated by the current episode of
star formation (these would tend to be galaxies with very
young inferred ages) could be more massive by factors as
large as $\sim 5-10$.
The typical BX/MD galaxy, which is
between these extremes, probably has a true stellar mass
that is a factor of $\leq 3$ larger than the value
inferred from population synthesis modeling.
Finally, this decomposition of the SED into
young and maximally-old components provides a fairly
robust (but extreme) upper limit to the stellar mass consistent with
the observed photometry, since more realistic, longer
current star-formation episodes, would make an increasing
contribution to the rest-frame near-IR light, and more
realistic past bursts would occur at $z < \infty$ and
have smaller $M/L$ ratios.

\begin{figure}
%\plotone{R_vs_m.eps}
\plotone{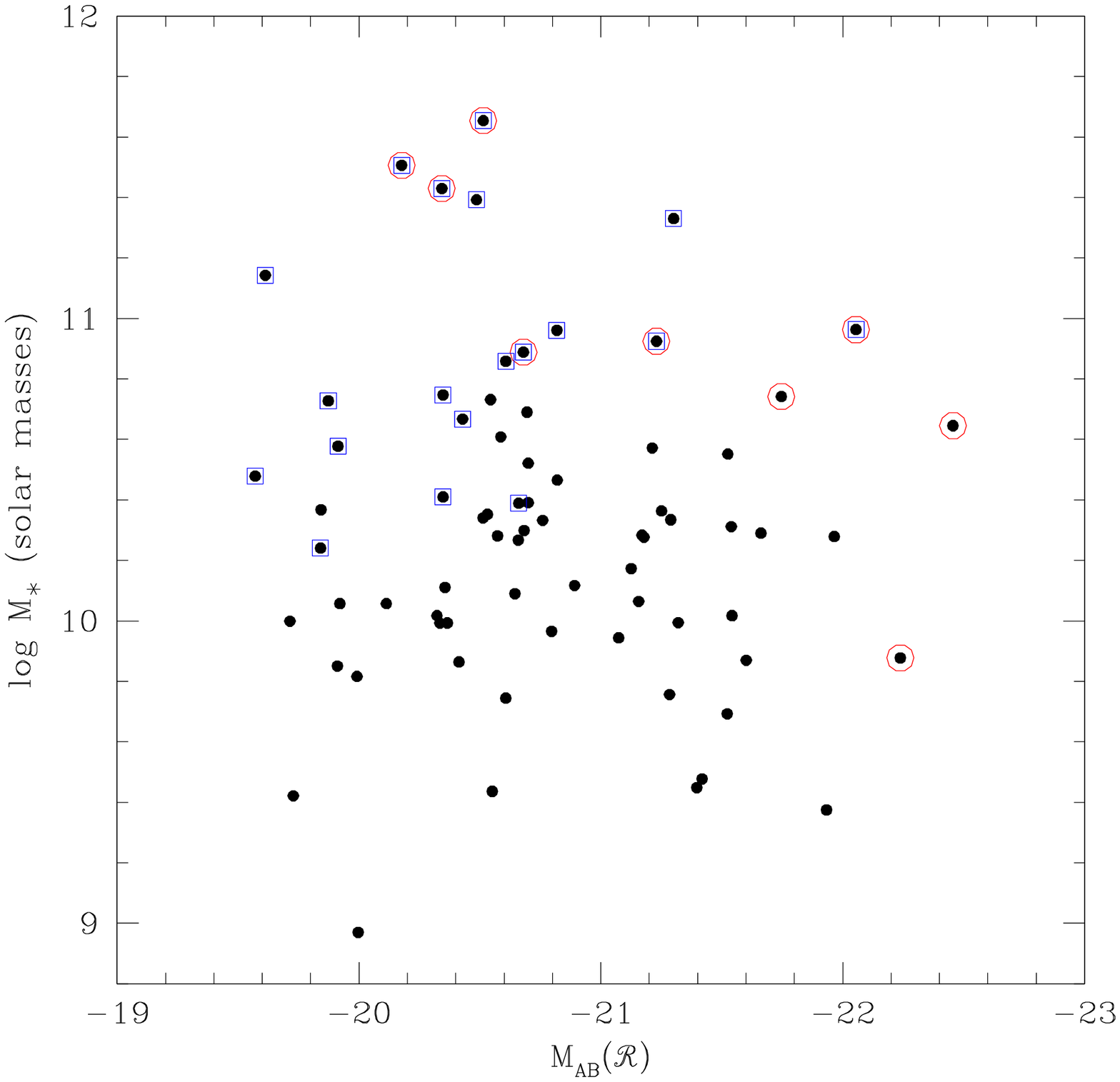}
\caption{The relationship between the absolute magnitude
in the observed ${\cal R}$ band (equivalent to rest-frame
2100 \AA\ at the mean redshift of the sample) and
inferred stellar mass. A Spearman test indicates no
correlation of these two quantities. There is a similar
lack of correlation between $M_{\ast}$ and the absolute
magnitude at rest-frame $\sim 1500$ \AA.
Again, objects with $K_s \le 20.0$ are circled and those
with ${\cal R}-K_s>3.5$ are indicated with open squares.}
\label{fig:R_vs_m}
\end{figure}

\subsection{AGN}
\label{sec:resultsagn}

As described above, there is one object, MD157,
whose spectrum shows clear signatures of AGN activity in
the far-UV (strong C~IV and He~II emission as well
as a broad Lyman $\alpha$ line). 
In addition, MD157
is extremely bright in the IRAC-4 band (AB mag of
18.51; see Table 2), such that obtaining a reasonable fit
to its SED using stellar populations required excluding
the 8$\mu$m data point.  Two other objects, MD174 and
MD94, have rest-frame optical spectra indicative of AGN
(both have H$\alpha$/[NII]$\simeq 1$, and broad
H$\alpha$ with $v_{FWHM} \geq 1000$~\kms). 
All three objects exhibit significantly
redder than average ${\cal R}-K_s$ colors, ranging from
$4.46 - 5.15$ and best-fit stellar masses ranging from
$8\times 10^{10} M_{\odot}$ to $4.5 \times 10^{11}
M_{\odot}$. 

An important question is how much the
presence of the AGN affects the stellar mass estimates.
The large stellar masses are inferred because of the
bright rest-frame near-IR photometry and the very red
rest-frame UV/optical and UV/near-IR colors, but if a
dusty AGN were dominating the rest-IR part of the SED,
these stellar mass inferences could be erroneous. A large
fraction of other similarly bright and red $z \simgt 2$
galaxies having relevant spectroscopic measurements also
indicate the presence of AGN. For example, 4 of the 7
galaxies with spectroscopic redshifts from the FIRES
survey \citep{franx2003} presented by \citet{vandokkum2003,vandokkum2004}
exhibit UV and/or optical signatures of AGN
activity (high ionization UV lines, broad H$\alpha$ and
H$\alpha$/[NII] line ratios near unity), similar to
MD157, MD174, and MD94.
It is difficult to know whether AGN activity leads to
erroneously large stellar mass estimates, or whether it
is preferentially objects that have already accumulated
large stellar masses that tend to harbor significant AGN
activity. When the AGN is not obviously energetically
dominant in the observed-frame mid-IR, we suspect that
the stellar mass estimates are probably reasonable; for
objects like MD157, which is obvious as an AGN in both
the rest-frame far-UV and near-IR, it is probably safest
to assume that the stellar mass has been over-estimated
using SED fitting.

There are other objects that appear to have fluxes at
8$\mu$m that are significantly in excess of that expected
from the SED modeling (e.g., BX756, BX535, MD69, BX505,
BX581, BX561, BX523) but where we do not have any obvious
spectroscopic evidence for the presence of an AGN. It is
conceivable that these mid-IR excesses could be due to
obscured AGN activity; deep Chandra images of this field
could help resolve this question.

\section{Discussion}
\label{sec:discussion}

We have seen above that modeling of the broad-band SEDs
of $z \sim 2$ star forming galaxies has significant
limitations, even with spectroscopic redshifts and when
longer-wavelength data are available for direct
measurement of the rest-frame near-IR light. We have
shown that reasonably consistent stellar mass estimates
are obtained independent of the exact form of the star
formation history, but that this consistency should be
viewed with caution because of the fact that a robust
episode of star formation can change the inferred $M/L$
by a factor of several, even at rest-frame $\sim 1.5$ $\mu$m.
One obtains increasingly reliable estimates of the total
stellar mass of a galaxy as the integral of star formation
over the lifetime of the galaxy exceeds that in
the past several hundred Myr by a factor of a few.  
In the most reliable cases, where the UV light is far surpassed
by the emission at rest-frame optical and near-IR wavelengths,
the inferred stellar mass does not depend
very strongly on whether a smooth or episodic prescription
is used to model the way in which the integrated stellar mass
accumulated as a function of time.
When the rest-frame UV luminosity is more comparable to rest-frame optical
and near-IR luminosities, however, it is inevitable that the stellar
masses inferred from photometric techniques will reflect
the typical stellar mass produced in an episode of star
formation, perhaps caused by a merger or accretion event
over the course of the last few hundred Myr, and not the
total stellar mass. 
A corollary to the above arguments is that it is most
likely one would measure large stellar masses (using the
type of modeling described in this paper) when the
current rate of star formation is significantly smaller
than the past average. Otherwise, most of the rest-frame
near-IR and optical light will be ascribed to the current
star-formation episode, which has much smaller $M/L$ even
in the IRAC bands. 
These are not new arguments \citep[see, e.g.,][]{papovich2001,shapley2001},
though the inclusion of IRAC wavelengths to constrain
the rest-frame near-IR emission places them on a more
robust footing, and the larger range of galaxy
properties observed in the current sample
highlights the important trends more clearly.

The concept of episodic star formation
informs the above discussion. While such variable
star-formation histories appear
a reasonable representation of the build-up of
stellar mass in galaxies at high redshift,
simply based on general consideration of the
hierarchical model of galaxy formation
in which merging (and merger-induced star formation)
plays a large role, we also find direct evidence from the 
galaxies in our sample. Specifically, we can
obtain interesting constraints on the redshift
at which the onset of star formation occurred for
the galaxies in our sample, despite large
systematic uncertainties in the stellar population
ages (section~\ref{sec:modsysunc}).
The maximum stellar population age
is obtained for a given set of rest-frame
UV to near-IR colors when a continuous star-formation
history is used, so we can use the CSF results 
to infer the earliest redshifts at which the current
episodes of star formation could begin \citep[cf.,][]{ferguson2002}.
The median CSF age for the sample is 
$\sim 600$~Myr. Using the redshift and CSF age distributions
we find that the current episode of star formation had not
begun yet at $z=3$ for at least $\sim 65$\% (46/72) of the galaxies
in the sample. The true fraction is probably higher,
since CSF models provide only an upper limit on the age
of the stellar population dominating the current
star-formation episode. Therefore, we find 
that the majority of galaxies in our sample
have young ages for the current
episode of star formation compared
to the age of the Universe at $z\sim 2$, allowing the possibility
of much older previous bursts of star formation. Relatively young inferred
ages would naturally arise in a picture of episodic star-formation.

A related point is that selecting on the basis of far-UV
properties, as we have done for the BX/MD sample,
connects only obliquely to stellar mass. One might expect
that galaxies would have to attain a particular total
mass threshold before they could support star formation
detectable at the apparent magnitude limit of the survey,
but once that threshold is exceeded, a combination of
extinction (which is probably star-formation rate
dependent; e.g., \citet{adelberger2000}) and scatter in
SFR per unit total (or stellar) mass might ``randomize''
the mix of galaxy properties included in the sample.
Indeed, as shown in figure~\ref{fig:R_vs_m}, we find that
there is no hint of a correlation between $M_{UV}$ and
inferred stellar mass, as was found previously in the $z
\sim 3$ LBG sample of \citet{shapley2001}.

These issues make it particularly difficult to measure
the total stellar mass that has formed by a particular
redshift \citep{dickinson2003,
glazebrook2004,fontana2004}, or to make a detailed
comparison of the stellar mass of even similarly-selected
galaxies at different redshifts. 
For example, analysis of the spatial clustering
of $z\sim 3$ LBGs and $z\sim 2$ BX galaxies \citep{adelberger2004b}
suggests that correlation length of LBG descendants at $z\sim 2$
would be similar to the observed BX correlation length,
with the implication that $z\sim 2$ LBG descendants reside
in dark matter halos in the same mass range as those
hosting BX galaxies. It is therefore of interest also 
to compare the stellar mass distribution of
spectroscopically-confirmed BX/MD objects 
with that of $z \sim 3$ LBGs from the sample of
\citet{shapley2001}. A direct evolutionary 
comparison between the stellar mass distributions for
$z\sim 2$ and $z\sim 3$ UV-selected samples is not
straightforward for many reasons. 
Most importantly, while there is little doubt that the objects at 
$z\sim 2$ and $z\sim 3$ with large inferred masses and ages really are
massive and old (see section~\ref{sec:discussmassive}),
the modeling results likely represent lower limits on the
stellar masses for more typical objects in both samples, where 
the current star formation is producing enough optical and
near-IR light to potentially mask previous generations of stars.
If star formation is episodic, as seems
quite likely on many grounds, then it seems possible in
principle that the same galaxy could be measured at
different times, but the {\it inferred} stellar mass
might not even be a monotonically increasing function of
time.  Similarly, it might be difficult to trace the
evolution of stellar mass during the epoch when most
galaxies are still very actively star-forming.

Bearing all of these concerns in mind, 
we now summarize the inferred properties
of typical UV-selected galaxies at $z\sim 2$. We then turn in some
detail to the most massive and oldest galaxies in the sample,
for which we obtain robust total stellar mass measurements,
and show how these massive galaxies provide plausible
progenitors for passive galaxies identified at $z\sim 1.5-2.0$.
Finally, we make a general comment about the relationships
among galaxy populations selected with different photometric criteria.

\subsection{The Properties of the ``Typical'' UV-Selected
Galaxy at $z \sim 2$}
\label{sec:discusstypical}

The results of this paper suggest
that the typical $z \sim 2.2$ ``BX'' galaxy has a stellar
mass of $\sim 3 \times 10^{10}$ M$_{\sun}$, a number
that could possibly be higher by a factor of $\sim 2-3$
without affecting the observed SEDs.  Although the length
of time stars have been forming in individual galaxies is not
well-determined by the modeling, 
 \citet{steidel2005} show that galaxies within the
significant over-density at $z=2.30$ are older
and contain more stellar mass on average than those
in lower-density environments. It is very clear from
these results that even at $z
\sim 2$ the star-formation history is strongly dependent
on large scale environment-- more so than on the
particular redshift at which the galaxies are observed.
The typical current star-formation rate is $\sim 50$
M\subsun yr$^{-1}$ \citep{reddy2004}, although unobscured
star formation down to $\sim 3$ M\subsun yr$^{-1}$ could
make it into the sample, and we know that among the BX
sample lurk sub-mm sources that have implied SFR of up to
$\simgt 1000$ M\subsun yr$^{-1}$ \citep{chapman2004}.
UV-selected galaxies that are also optically bright ($K_s
\simlt 20$) have typical inferred SFR higher by a factor
of $\sim 3$ \citep{reddy2005a}. 

The space density of
galaxies similar to those discussed in this paper is
$\sim 3\times10^{-3}$ Mpc$^{-3}$ at $z\simeq 2.2$
\citep{reddy2005b}. The co-moving correlation length of
the UV-selected galaxies at $z \sim 2$ is $r_0=4.2\pm0.4$
h$^{-1}$Mpc \citep{adelberger2005}, implying that the
typical galaxy lives in a dark matter halo of mass $\sim
2\times10^{12}$ M\subsun. Assuming that the galaxies have
a baryon fraction that is the same as the universal
value, the estimated mass in baryons associated with the
halo is $\sim 3\times10^{11}$ M\subsun, suggesting that
on average roughly 10\% of the baryons have been
converted into stars.  Consideration of the expected
evolution of the clustering properties of the UV-selected
galaxies suggests that by $z \sim 1$ they will be found
mainly among early-type galaxies, and will have completed
most of their star formation. The most massive examples
will have completed their star formation even earlier,
and exhibit significantly stronger clustering than the
average UV-selected galaxies at $z \sim 2$
\citep{adelberger2005}.
As discussed further in section~\ref{sec:discussmassive}
and other recent work,
the $\sim 2.5$ Gyr between $z \sim 2$ and $z
\sim 1$ marks the transition of a large fraction of
massive galaxies from strongly star forming to largely
quiescent \citep{daddi2004b,shapley2004,cimatti2004,mccarthy2004,fontana2004}.

\subsection{Massive Galaxies in Rest-UV-Selected Samples}
\label{sec:discussmassive}

The most massive galaxies in the current sample, with
stellar masses $> 10^{11} M_{\odot}$, cannot contain
significant additional stellar mass that is masked
by the current episode of star formation. We obtain
roughly the same stellar mass estimate regardless of
whether a single-component model is used to fit
the colors, or a young current star-formation
episode super-imposed on an old past burst. 
In the context of simple, single-component models,
these galaxies also have SEDs that
provide significant constraints on the star-formation
history. Unlike the majority of galaxies in the HS1700
sample, for which we find that all or most $\tau$ values
are consistent at the 95.4\% level with the data, we can
rule out models with $\tau < 200$~Myr and $\tau=\infty$
(i.e., CSF)  for all but one of the galaxies with
$M_{\ast} > 10^{11} M_{\odot}$. More discrimination among
$\tau$ values leads to interesting constraints on the
duration of the star-formation episodes in these
galaxies. The minimum $\tau=200$~Myr for these five
galaxies implies minimum ages of $t_{sf} \geq
600-900$~Myr. The most likely value of $\tau$ however is
$1-2$~Gyr, based on Monte Carlo simulations in which
$\tau$ was allowed to vary. For $\tau=1-2$~Gyr, the ages
inferred for these galaxies range from $2-3$~Gyr,
comparable to the age of the Universe at $z\sim 2.3$. The
stellar masses for this subsample of galaxies are all
well-constrained ($\leq 20$\% uncertainties), including
the systematic uncertainty in star-formation
history\footnote{ We note that the massive galaxies in
the current sample are quite similar to those presented
by \citet{shapley2004}, which were all drawn from another
of our survey fields, Q1623+26.}.

The superior constraint on star-formation history for
these objects results from two factors: first, the most
massive galaxies are among the brightest in the $K_s$ and
IRAC bands\footnote{We note, for comparison with the K20
survey \citep{cimatti2002} that only half of the galaxies
in our sample with $M_{\ast} > 10^{11} M_{\odot}$ also
have $K_s \leq 20.0$.  A deeper apparent magnitude limit
than $K_s=20.0$ is evidently necessary to gain a complete
census of the most massive galaxies at $z \sim 2$
\citep{glazebrook2004}.}, resulting in smaller than
average ${\cal R}-K_s$ and IRAC photometric
uncertainties. Four of the five massive galaxies with
significant constraints on $\tau$ are also detected in
all four IRAC bands, providing better constraints on the
shape of the SED in the region dominated by longer-lived
stars.  Secondly, all but one of the objects with
$M_{\ast} > 10^{11} M_{\odot}$ have significantly redder
than average ${\cal R}-K_s$ (and ${\cal R}-IRAC$) colors\footnote{Most
of the galaxies with $M_{\ast} > 10^{11} M_{\odot}$ have best-fit models
with $J-K_s>2.3$, implying that they would be
selected by the FIRES survey.},
but average or bluer than average $G-{\cal R}$ colors.
The combination of blue to average $G-{\cal R}$ and red
${\cal R}-K_s$ colors indicates
that the rest-frame optical and near-IR light
cannot be ascribed to the massive stars tracing the
instantaneous star-formation rate and dominating the
rest-frame UV emission, and therefore
must be produced by less massive
stars formed primarily in the past. Since the inferred
extinction correction will be small when the UV color is
blue, either the star formation was much higher in the
past, or the current SFR has been sustained for a very
long time. If the implied CSF time is older than the age
of the universe, then $\tau$ models will be required to
fit the data well.

There are other objects in the sample with comparably
small ${\cal R}-K_s$ and IRAC uncertainties, but with
weaker constraints on $\tau$. An example is MD94, which
has very red ${\cal R}-K_s$ and $G-{\cal R}$ colors and
is inferred to have a stellar mass $M>10^{11} M_{\odot}$.
In this case, the colors can be adequately fit by models
spanning the whole range of $\tau$, including CSF, mainly
because the inferred current SFR (derived from the
extremely red $G-{\cal R}$ color) is high enough to have
produced the near-IR light in much less than the age of
the universe, even for CSF. The best-fit $\tau$ model,
though, has an age that is a factor of $\sim 3$ shorter
than the CSF model, and a current SFR 3 times smaller
(however, the implied initial SFR given the age and
$\tau$ values would have been 25 times larger, or 2500
M$_{\sun}$ yr$^{-1}$!).

We return now to the massive galaxies with superior constraints
on the single-component star-formation histories.
Of course, a smoothly-declining star-formation history is
not a very likely scenario for the build-up of stellar
mass, since the star-formation rate
is expected to be modulated by episodic merger and accretion
events. It is not really possible to establish whether or
not star formation has been sustained continuously at a
detectable level over a long period of time, even for
galaxies whose star formation must have begun at very
high redshift. For example, in general, galaxies for
which the $\tau$ models are required for a good fit to
single-component models can also be adequately fit by
assuming a single, early burst of star formation to
produce the optical/near-IR light and a small amount of
recent star formation (with an associated stellar mass of
only $\sim 1-2$\% of the early burst) that accounts for
the rest-frame far-UV light.  For such two-component
models, there is again no unique solution, but
reasonable fits to the data can be found for models where
a large burst with $\tau = 100$Myr occurred at $z > 4$
and the recent (smaller) burst is less than 100 Myr
old\footnote{We note that such models would imply SFRs
that averaged $\sim 1500$ M$_{\sun}$ yr$^{-1}$ over the
first 200 Myr of the galaxy's life, e.g. between $z=5$
and $z=4.4$.}.  While we cannot distinguish between
smooth or episodic star-formation histories, 
the estimates of total stellar mass derived from two-component
models agree very well with those obtained from the
simple single-component fits. For these massive and old
galaxies, the inferred total stellar mass is therefore a robust
quantity.

In an analysis of the stellar populations of UV-selected
$2.0 \leq z \leq 3.5$ galaxies in the {\it Hubble Deep
Field} (HDF-N) \citet{papovich2001} found no comparably
massive and old objects. The small ratio between the
typical inferred star-formation timescale for these HDF
LBGs and the time interval between $2.0 \leq z \leq 3.5$
was used to motivate a scenario of multiple, intermittent
star-formation episodes with a duty cycle of $\leq
1$~Gyr. As described in the beginning of section~\ref{sec:discussion}, 
most of the current sample can in fact be described in this manner.
In retrospect, it is not surprising that the
sample of \citet{papovich2001} did not contain any
$M_{\ast} > 10^{11}$ M$_{\sun}$ galaxies in the 5.0
arcmin$^2$ HDF-N field, despite a similar optical
magnitude limit in the HDF spectroscopic sub-sample.  We
found only 6 with spectroscopic redshifts in the entire
65 arcmin$^2$ WIRC field. Since the spectroscopic
completeness of our sample is $\sim 22$\%, the
approximate surface density of objects in the redshift
interval $z \simeq 2.3\pm0.4$ is $\sim 0.5$
arcmin$^{-2}$, so that one would expect $<2$ such massive
and old galaxies in an area with the size of the HDF-N
given spectroscopic incompleteness.  In fact,
\citet{labbe2003} have pointed out the discrepancy in the
number of galaxies with red near-IR colors in the HDF-N
versus their FIRES field in the HDF-S, with the HDF-N
having only 10-20\% of the numbers in the HDF-S.  Given
that the galaxies with large stellar mass are strongly
clustered (\citet{daddi2004}, \citet{adelberger2005}) the
variance between small survey areas is expected to be
large for such rare objects.

\subsection{UV-Selected Progenitors of Massive and
Passive Galaxies?}
\label{sec:discusspassive}

While massive and mature galaxies represent a small
fraction by number of the BX/MD population (6/72 with
$M\geq 10^{11} M_{\odot}$, 5 of which also have
$t_{SF}\sim $~the age of the Universe), their estimated
space density, $\sim 10^{-4} \mbox{Mpc}^{-3}$, is similar
to the space density of the massive star-forming galaxies
from the K20 survey presented by \citet{daddi2004} at
$1.7 \leq z \leq 2.3$, and also the massive galaxies with
old and passive stellar populations presented by
\citet{cimatti2004} and \citet{mccarthy2004} at $1.3 \leq
z \leq 2.0$. In both of the latter papers, the authors
suggest that the progenitors of the old and passive
galaxies are to be found among submillimeter- and
near-IR-selected samples at $z>2$. Here we show that the
most massive and mature objects in the $2.0 \leq z \leq
2.6$ UV-selected sample could quite plausibly be the
progenitors of massive and passive systems at $1.3 \leq z
\leq 2.0$. To make this comparison, we adopt star
formation histories that are consistent both with those
used to fit the near-IR selected massive and passive
galaxies by \citet{cimatti2004} and \citet{mccarthy2004}
and with the SEDs of all 5 of the BX/MD massive and old
galaxies at $z \sim 2.3$.
We then evolve these models forward in time to the mean
redshifts of the \citet{cimatti2004} and
\citet{mccarthy2004} samples, compute the average $K_s$
magnitudes, $R-K_s$ and $I-K_s$ colors, stellar masses,
and star-formation rates, and compare them with the
properties of the passive galaxies.

\citet{cimatti2004} present four galaxies at $\langle z
\rangle =1.7$, with $\langle K_s \rangle = 18.7$,
$\langle R - K_s \rangle = 6.3$, stellar masses ranging
from $1-3 \times 10^{11} M_{\odot}$, and a space density
of $\sim 10^{-4}\mbox{Mpc}^{-3}$. When a declining model
with $\tau=0.3$~Gyr is used to fit the colors, the
associated ages and formation redshifts are $\sim 2$~Gyr,
and $z\sim 4$, respectively. If we apply a $\tau=0.3$Gyr
model to the massive BX/MD SEDs at $\langle z \rangle =
2.3$, and then evolve them forward to $z=1.70$, they
would have $\langle K_s \rangle = 19.4$, $\langle R - K_s
\rangle = 6.2$, $\langle M* \rangle = 2.2\times 10^{11}
M_{\odot}$, ages ranging from 1.9-2.3 Gyr, and SFRs of
$\le 1$ $M_{\odot}\mbox{yr}^{-1}$, making them fainter
than ${\cal R}=25.0$.  We find a similar level of
agreement between the BX/MD models evolved to $z=1.5$ and
the parameters of the 5 objects in \citet{mccarthy2004}
(their table 1) for which a $\tau=0.5$~Gyr model was used
to fit the spectra and colors.  Thus, extrapolating star
formation histories for the massive BX/MDs forward in
time would make them plausibly consistent with the
passive galaxies in the near-IR selected samples observed
at somewhat lower redshift. Equivalently, if the BX/MD
examples are interpreted as massive old bursts with a
small amount of recent star formation superposed, by $z
\sim 1.7$ they would have essentially the same
characteristics as the passive galaxies provided that the
current episode of star formation (i.e., as observed at
$z\sim 2.3$)  lasts for less than the $\sim 900$~Myr
between $z \sim 2.3$ and $z \sim 1.7$.

The specific values of the timescale over which
star formation occurs, and the redshift at which
star formation begins, are not well-determined because of
the systematic uncertainties inherent in stellar
population synthesis modeling. However, we have
demonstrated that, for reasonable assumptions (or at
least consistent with those used by other workers), the
most massive and evolved galaxies in the UV-selected
sample are likely to evolve into largely passive systems
by $z\sim 1.7$, with masses, ages, inferred formation
redshifts, colors, and space densities similar to those
of the massive and old stellar populations presented in
the K20 and GDDS surveys. While additional progenitors of
the $1.3 \leq z \leq 2.0$ massive and passive systems may
be found among submillimeter- and $J-K$ color-selected
samples, the UV-selected sample at $z \sim 2.3$ must
contain a significant fraction of the objects that will
look like elliptical galaxies at $1.3 \leq z \leq 2.0$.

\citet{adelberger2004b} have reached similar conclusions
about the fate of rest-UV-selected galaxies, using
independent lines of argument. On the basis of a
comparison of the observed clustering properties and
galaxy number density at $z \sim 2$ with dark matter
halos in numerical simulations, Adelberger \et argue that
the BX galaxies are consistent with typical total halo
masses of $\simgt 10^{12}$ M$_{\sun}$ and that by $z\sim
1$ they will have a clustering strength similar to that
of galaxies with early type spectra in the DEEP2 redshift
survey results of \citet{coil2004}. In this context, the
most massive BX galaxies at $z \sim 2$ would be among the
more extreme (in terms of clustering properties) early
type galaxies by $z \sim 1$.  \citet{adelberger2005} has
also shown that optically bright ($K_s \le 20.5$)  
rest-UV-selected galaxies at $z \sim 2$, of which the
most massive galaxies in the present sample are a subset,
are as strongly clustered as their rest-optically
selected counterparts \citep[cf.,][]{daddi2003,daddi2004}.

\subsection{Selection Biases and ``Populations''}
\label{sec:discussbias}

We consider it likely that real galaxies actually evolve
in and out of high redshift galaxy samples selected in
various ways.  For example, the massive galaxies in the
BX/MD sample could easily have been sub-mm galaxies at $z
\sim 4$ (when some versions of their star-formation
histories predict high enough SFRs for them to have been
easily detectable at 850 $\mu$m, although most would
probably not have been $\mu$Jy radio sources because
their redshift is too high), and may have oscillated
between being quiescent and being UV-selectable depending
on whether or not there was a trace of relatively
unobscured star formation superposed on the more mature
stellar populations. 

As discussed by \citet{shapley2004} and
\citet{adelberger2004b,adelberger2005}, the most massive and optically
brightest galaxies in the UV-selected samples appear to
have similar masses, chemical abundances, star-formation
rates, inferred ages, and clustering properties as
galaxies selected in the near-IR by either apparent
brightness or optical/IR color.  
This comparison shows that a UV-selected survey
is capable of finding examples of galaxies
that are very similar to those selected in other ways;
it is not a demonstration that a UV-selected survey
``finds'' the vast majority of galaxies at $z\sim 2$. 
That rest-UV selection is incomplete with
respect to all galaxies present at similar redshifts has
been shown clearly by the near-IR selected GDDS, K20, and
FIRES surveys as discussed above, as well as by sub-mm
selected surveys (e.g., \citet{chapman2003}).  The point
is that it is relatively easy for objects to come and go
from a UV-selected surveys on timescales much shorter
than a Hubble time at the relevant redshifts, independent
of their past (or future) star-formation histories. Most
of the existing evidence, and some theoretical studies
(e.g., \citet{nagamine2004}), suggests that at $z \sim
2$, relatively massive galaxies ($M_{\ast} \simgt
10^{10}$ M$_{\sun}$) are ``UV-selectable'' using criteria
similar to those employed in the current sample $\simgt
50$\% of the time. 

While such samples would be
significantly incomplete with respect to all galaxies at
a given stellar mass, they may not be entirely missing
any type of galaxy. The fraction that happens to have
enough UV light to be selected has significant
advantages: spectroscopy is easy in both the far-UV and
optical (rest frame). These observations allow for more detailed
astrophysical studies, and an evaluation of the
dependence of galaxy properties on large-scale
environment as in \citet{steidel2005}. The recent
success of the spectroscopic follow-up of sub-mm-selected
galaxies \citep{chapman2004} has benefited from exactly
the same advantage: even UV light that is inconsequential
compared to the bolometric luminosity allows
spectroscopic measurements, leading to more detailed
astrophysical insight.

\section{SUMMARY}
\label{sec:summary}

We have presented a detailed analysis of the broad-band SEDs
of a sample of 72 rest-UV selected, spectroscopically
confirmed, star forming galaxies at $z=2.3\pm0.3$ in a
$\sim 8.5$ by 8.5 arcmin field centered on HS1700+643.
This sample of galaxies is broadly representative of
galaxies selected with well-calibrated photometric
pre-selection using rest-UV colors: all of the
spectroscopic data were obtained without consideration of
the longer-wavelength properties.  We used extensive
ground-based photometry from 0.35-2.15 $\mu$m and deep
{\it Spitzer}/IRAC data probing the
rest-frame near-IR, together with 
information from optical and near-IR spectroscopy, to
produce an unprecedented data set that allows the most
robust statements to date concerning the physical state
of galaxies at $z \sim 2$, in particular their stellar
masses. 
 
Our main conclusions include the following:

1. Using population synthesis to model the stellar
populations, we find that the inferred stellar mass
distribution of the sample is reasonably well-described
by a log-normal distribution: $\langle log {M_{\ast}
\over M_{\sun}} \rangle = 10.32 \pm 0.51$ and that the
result is essentially unchanged if a constant star
formation history is assumed instead of the best-fitting
exponentially declining star-formation history.  The
typical uncertainty in stellar mass, including variations
in the assumed form of the star-formation history, is
$\sim 30-35$\%. Sub-samples of the galaxies with brighter
observed $K_s$ (2.15$\mu$m) flux have larger inferred
stellar mass; 8\% of the sample has inferred stellar mass
$> 10^{11}$ M\subsun.  We found that fitting SEDs to
simultaneously determine photometric redshifts and
stellar population parameters leads to large
uncertainties in both, and tends to over-estimate the
inferred stellar mass; thus we use only the sample with
spectroscopic identifications.

2. The use of the {\it Spitzer}/IRAC fluxes, which
measure the rest-frame near-IR light from the galaxies
and thus should yield the strongest available
constraints on stellar mass, provides a marked improvement
in the estimates of the stellar mass of individual
galaxies, reducing the formal uncertainties by a factor
of $1.5-2$ relative to estimates based on photometry that
extends only to the ground-based $K_s$ band. However, we
find that the longer wavelength data do not change the
results significantly: the total stellar mass inferred
for the entire sample of 72 galaxies is changed by only
15\% depending on whether the {\it Spitzer} data are used for
the modeling or not. At the redshifts spanned by our
sample, there is a tendency to over-estimate slightly the
stellar mass when the $K_s$ band is the longest
wavelength observed, because of contamination
in the $K_s$ band from $\Ha$ emission.

3. The lack of strong constraints 
on the form of the star-formation
history remains a problem for most of the galaxies
in the sample, even when using
simple, single-component models. Ages, exponential time
constants, and extinction/SFR are largely degenerate with
respect to each other; however, as found in previous work,
the inferred stellar masses remain largely unaffected. External
constraints can be used to rule out
parts of parameter space that are otherwise allowed by
the SED fits.

4. While the inferred stellar mass is correlated with the
flux at 2 $\mu$m, the inferred stellar mass-to-light
ratios can vary by a factor of $\sim 70$ at a given
$K_s$-band luminosity (rest-frame R at the mean redshift of
the sample).  Even at an observed wavelength of 4.5$\mu$m
(rest-frame 1.4 $\mu$m) the inferred $M/L$ varies by a
factor of 15. At both wavelengths (rest-frame 0.65 and 1.4
$\mu$m), the highest inferred $M/L$ are similar to those of
local galaxies, but the average $M/L$ is a factor of at least 5 and
2.5 smaller, respectively. Thus, at these redshifts, even
rest near-IR luminosities are by themselves a relatively
poor proxy for stellar mass; the entire SED, from the UV
to the near-IR, is essential. In particular, the
conversion from rest-frame near-IR luminosity to stellar mass is
strongly correlated with the observed ${\cal R}-K_s$
color (rest-frame $2000 - 6500$ \AA) which in the context
of the models considered is a crude measure of the ratio
of past to current star formation.

5. Even with IRAC rest-frame near-IR information,
there remains the possibility of
significant systematic underestimates of the stellar mass
using single-component
SED-fitting techniques; the magnitude of this effect
depends sensitively on the shape of the SED. For objects
that are red in their UV/optical and UV/near-IR colors, the
mass inferred from SED fitting is unlikely to
underestimate the stellar mass by more than a few tens of
percent, while the bluest objects in the sample could
have total stellar masses that are as much as a
factor of $\sim 5-10$ larger than inferred
(in the case of the most extreme assumed star-formation histories). 
A typical object in our sample
could be ``hiding'' up to a factor of $\sim 3$ more
stellar mass than inferred by the modeling. Thus, for
star-forming galaxies at high redshift, reasonably
precise estimates of stellar mass are possible only for
objects that are nearly finished forming stars, or caught
in a relatively quiescent state between episodes of star
formation. All other estimates should probably be
considered as lower limits, possibly reflecting the
typical stellar mass formed during a star-formation
episode rather than the integrated stellar mass.

6. We have shown that the most massive galaxies in the
UV-selected sample, with $M_{\ast} > 10^{11}$ M\subsun,
could plausibly evolve by $z \sim 1.7$ into the high
redshift ``passive'' galaxies recently identified from
near-IR selected spectroscopic surveys. 
The most likely star-formation
histories for these UV-selected examples indicate that they
began forming stars at very early times and that their
star-formation rates were considerably higher in the
past, quite similar to what is found for the near-IR selected galaxies.  
Taken together with the similarities in space
density, stellar mass, chemical abundances, and
clustering properties, galaxies with similarly large
stellar masses should probably be treated as a single
``population'' regardless of the method used to discover
them.  In most cases, the UV light which brings them into
the sample represents a small fraction of the stellar
mass assembly.

This work is based in part on observations made with the Spitzer Space Telescope,
which is operated by the Jet Propulsion Laboratory, California Institute
of Technology under NASA contract 1407.  Support for this work was
provided by NASA through contract 125790, issued by JPL/Caltech.
We are indebted to the IRAC instrument team, particularly
Giovanni Fazio and Peter Eisenhardt, for
obtaining and reducing the {\it Spitzer}/IRAC data in the
HS1700+64 field, and for allowing us early access to the
data.  Kevin Bundy kindly provided excellent software for
the reduction of the WIRC data.  Paul Francis,
the referee, provided comments that improved the
manuscript. CCS, DKE, and NAR have
been supported by grant AST03-07263 from the US National
Science Foundation and by the David and Lucile Packard
Foundation.  KLA acknowledges support from the Carnegie
Institution of Washington, and AES from the Miller
Foundation for Basic Research in Science.
We wish to extend special thanks to those of Hawaiian ancestry on
whose sacred mountain we are privileged to be guests. Without their generous
hospitality, the spectroscopic observations presented herein would not
have been possible.

%\bibliographystyle{apj}
%\bibliography{apj-jour,lbgrefs}

\begin{deluxetable}{lcccccc}
\tabletypesize{\footnotesize}
\tablewidth{0pt}
\tablecaption{HS1700$+$64 Imaging Observations\label{tab:obs}}
\tablehead{
\colhead{Band} &  \colhead{$\lambda_{\rm eff}$($\mu$m)} & \colhead{Telescope\tablenotemark{a}} & \colhead{Field Size} & \colhead{Integration} & \colhead{Seeing/Image Size}  
& \colhead{Photometric Depth\tablenotemark{b}} 
} 
\startdata
${\rm U_n}$ & 0.36 & WHT/Keck I 2001 May & $15\minpoint3 \times 15\minpoint3$ & 23400/3600 & $0\secpoint89$ &  26.4 \\
${\rm G}$ & 0.47   & WHT 2001 May        & $15\minpoint3 \times 15 \minpoint3$ & ~9900 & $0\secpoint83$ &  26.5 \\ 
${\cal R}$ & 0.68   & WHT 2001 May        & $15\minpoint3 \times 15 \minpoint3$ & ~7800 & $0\secpoint78$ &  25.7 \\
$\rm K_s$ & 2.15  & P200 2003 June/Oct  & $8\minpoint5 \times 8\minpoint5$   & 39720 & $0\secpoint84$ &  24.0 \\ 
IRAC-1 &  3.6 & {\it Spitzer} 2003 Oct   &   $5\arcm \times 10\arcm$  & 39600\tablenotemark{c} & $2\secpoint02$ &  24.6 \\
IRAC-2 &  4.5 & {\it Spitzer} 2003 Oct   &   $5\arcm \times 10\arcm$  & 39600\tablenotemark{c} & $2\secpoint38$ &  24.3 \\
IRAC-3 &  5.8 & {\it Spitzer} 2003 Oct   &   $5\arcm \times 10\arcm$  & 39600\tablenotemark{c} & $2\secpoint36$ &  23.0 \\
IRAC-4 &  8.0 & {\it Spitzer} 2003 Oct   &   $5\arcm \times 10\arcm$  & 39600\tablenotemark{c} & $2\secpoint56$ &  22.8 \\

\enddata
\tablenotetext{a}{WHT: William Herschel 4.2m telescope with Prime Focus Imager; Keck I: Low Resolution Imaging
Spectrometer, blue side. P200: Palomar 5.1m telescope with Wide Field Infrared Camera; {\it Spitzer}: Spitzer Space
Telescope/Infrared Array Camera}
\tablenotetext {b}{AB magnitude for a 5$\sigma$ detection in a 2\arcs\ diameter aperture. For IRAC bands,
the magnitude for a 5$\sigma$ detection using PSF-fitting and local background estimation as described in
the text}.
\tablenotetext {c}{Total exposure time in the deepest part of the mosaic. In general, the deepest
regions of the IRAC-1 and IRAC-3 bands are in different parts of the field from those deepest
in IRAC-2 and IRAC-4.}
\end{deluxetable}

\clearpage
\LongTables
\begin{landscape}
\begin{deluxetable}{lccccccccccc}
%\rotate
\tabletypesize{\footnotesize}
\tablewidth{0pt}
\tablecaption{HS1700$+$64 Galaxies with Spectroscopic Redshifts\tablenotemark{a}\label{tab:phot}}
\tablehead{
\colhead{Name} &  \colhead{$\alpha(J2000)$} & \colhead{$\delta(J2000)$} & \colhead{z} & \colhead{${\cal R}_{\rm AB}$} &
\colhead{$(G-{\cal R})_{\rm AB}$}  & \colhead{$(U_n-G)_{\rm AB}$} & \colhead{${\cal R}-K_s$\tablenotemark{b}} & \colhead{(IRAC-1)$_{\rm AB}$}
& \colhead{(IRAC-2)$_{\rm AB}$} & \colhead{(IRAC-3)$_{\rm AB}$} & \colhead{(IRAC-4)$_{\rm AB}$}
}
\startdata
BX627 &17:00:34.435 & 64:11:29.699 & 1.478 & 24.44$\pm$0.14 & -0.10$\pm$0.05 & 0.12$\pm$0.05 & 2.36$\pm$0.51 & \nodata & 23.63$\pm$0.18 & \nodata & \nodata \\
BX738 &17:01:16.150 & 64:12:36.684 & 1.710 & 23.93$\pm$0.12 & 0.01$\pm$0.04 & 0.49$\pm$0.09 & 3.29$\pm$0.15 & 22.11$\pm$0.10 & 22.15$\pm$0.10 & 22.56$\pm$0.29 & 23.07$\pm$0.49 \\
BX756 &17:00:55.146 & 64:12:54.403 & 1.738 & 23.21$\pm$0.11 & 0.33$\pm$0.05 & 0.58$\pm$0.09 & 2.41$\pm$0.23 & \nodata & 21.82$\pm$0.10 & \nodata & 21.55$\pm$0.15 \\
BX681 &17:01:33.762 & 64:12:04.278 & 1.740 & 22.04$\pm$0.05 & 0.19$\pm$0.01 & 0.40$\pm$0.02 & 2.99$\pm$0.08 & 21.02$\pm$0.10 & 21.08$\pm$0.10 & 21.16$\pm$0.18 & 21.35$\pm$0.14 \\
BX843 &17:01:15.627 & 64:13:52.891 & 1.771 & 24.61$\pm$0.16 & 0.18$\pm$0.08 & 0.51$\pm$0.11 & 3.26$\pm$0.31 & 22.95$\pm$0.34 & 22.80$\pm$0.10 & \nodata & \nodata \\
BX470 &17:01:26.563 & 64:09:31.993 & 1.841 & 24.26$\pm$0.14 & 0.38$\pm$0.07 & 0.75$\pm$0.13 & 4.20$\pm$0.11 & 21.44$\pm$0.10 & 21.45$\pm$0.10 & 21.30$\pm$0.10 & 21.69$\pm$0.13 \\
BX846 &17:01:02.925 & 64:13:54.470 & 1.843 & 23.91$\pm$0.12 & 0.09$\pm$0.04 & 0.48$\pm$0.09 & 3.18$\pm$0.18 & \nodata & 22.13$\pm$0.10 & \nodata & 22.15$\pm$0.33 \\
BX574 &17:00:46.904 & 64:10:49.536 & 1.846 & 24.10$\pm$0.14 & 0.24$\pm$0.07 & 0.73$\pm$0.10 & 3.43$\pm$0.19 & 22.31$\pm$0.10 & 22.27$\pm$0.10 & 22.06$\pm$0.22 & 22.99$\pm$0.36 \\
BX1043 &17:01:17.328 & 64:15:29.170 & 1.847 & 23.54$\pm$0.12 & -0.01$\pm$0.05 & 0.53$\pm$0.09 & 2.34$\pm$0.42 & \nodata & 22.78$\pm$0.24 & \nodata & \nodata \\
BX1087 &17:00:59.047 & 64:15:55.195 & 1.871 & 23.10$\pm$0.11 & 0.12$\pm$0.04 & 0.59$\pm$0.08 & 2.85$\pm$0.15 & \nodata & 21.93$\pm$0.10 & \nodata & \nodata \\
BX1032 &17:00:52.827 & 64:15:23.792 & 1.883 & 23.76$\pm$0.12 & 0.18$\pm$0.04 & 0.62$\pm$0.06 & 3.04$\pm$0.18 & \nodata & 22.37$\pm$0.10 & \nodata & 22.59$\pm$0.30 \\
BX782 &17:01:11.146 & 64:13:14.360 & 1.930 & 23.94$\pm$0.12 & 0.05$\pm$0.04 & 0.59$\pm$0.09 & 2.97$\pm$0.22 & 22.18$\pm$0.15 & 22.35$\pm$0.11 & \nodata & \nodata \\
BX530 &17:00:36.859 & 64:10:17.376 & 1.942 & 23.05$\pm$0.11 & 0.21$\pm$0.05 & 0.69$\pm$0.05 & 2.80$\pm$0.15 & 21.73$\pm$0.10 & 21.88$\pm$0.10 & \nodata & 22.19$\pm$0.33 \\
BX807 &17:00:54.909 & 64:13:28.593 & 1.971 & 24.83$\pm$0.16 & 0.08$\pm$0.08 & 0.45$\pm$0.11 & 3.04$\pm$0.36 & \nodata & 23.49$\pm$0.10 & \nodata & \nodata \\
BX903 &17:00:54.431 & 64:14:14.761 & 1.976 & 24.39$\pm$0.14 & 0.18$\pm$0.06 & 0.50$\pm$0.11 & 2.99$\pm$0.28 & \nodata & 22.80$\pm$0.10 & \nodata & \nodata \\
BX536 &17:01:08.939 & 64:10:24.949 & 1.977 & 23.00$\pm$0.11 & 0.21$\pm$0.05 & 0.79$\pm$0.05 & 3.27$\pm$0.10 & 21.23$\pm$0.10 & 21.20$\pm$0.10 & 21.23$\pm$0.10 & 21.83$\pm$0.12 \\
BX412 &17:01:20.109 & 64:08:37.363 & 1.981 & 23.58$\pm$0.12 & 0.31$\pm$0.06 & 0.63$\pm$0.06 & 3.31$\pm$0.15 & 22.13$\pm$0.10 & 22.23$\pm$0.10 & 22.32$\pm$0.14 & \nodata \\
BX526 &17:00:58.040 & 64:10:15.947 & 2.018 & 23.99$\pm$0.12 & 0.00$\pm$0.04 & 0.45$\pm$0.09 & 2.25$\pm$0.42 & 22.90$\pm$0.24 & 22.94$\pm$0.23 & 23.05$\pm$0.34 & \nodata \\
BX1075 &17:00:51.215 & 64:15:47.515 & 2.061 & 24.83$\pm$0.16 & 0.30$\pm$0.09 & 0.89$\pm$0.18 & 2.92$\pm$0.54 & \nodata & 24.46$\pm$0.48 & \nodata & \nodata \\
BX625 &17:01:11.493 & 64:11:29.242 & 2.078 & 24.52$\pm$0.16 & -0.03$\pm$0.07 & 0.49$\pm$0.06 & 2.96$\pm$0.54 & 23.28$\pm$0.10 & 23.28$\pm$0.10 & \nodata & \nodata \\
BX592 &17:01:02.703 & 64:11:03.422 & 2.099 & 24.87$\pm$0.16 & 0.38$\pm$0.09 & 1.11$\pm$0.19 & 3.01$\pm$0.36 & 23.35$\pm$0.11 & 23.43$\pm$0.12 & \nodata & \nodata \\
BX535 &17:01:07.597 & 64:10:26.403 & 2.113 & 25.16$\pm$0.16 & 0.57$\pm$0.12 & 1.55$\pm$0.37 & 3.45$\pm$0.37 & 23.04$\pm$0.14 & 22.89$\pm$0.17 & 22.82$\pm$0.22 & 22.78$\pm$0.33 \\
BX1012 &17:00:47.578 & 64:15:11.701 & 2.115 & 23.72$\pm$0.12 & 0.08$\pm$0.04 & 0.53$\pm$0.09 & 2.77$\pm$0.26 & \nodata & 23.45$\pm$0.45 & \nodata & \nodata \\
BX691 &17:01:05.996 & 64:12:10.271 & 2.190 & 25.33$\pm$0.16 & 0.22$\pm$0.10 & 0.66$\pm$0.19 & 4.64$\pm$0.17 & 21.89$\pm$0.10 & 21.65$\pm$0.10 & 21.44$\pm$0.12 & 22.02$\pm$0.27 \\
BX1118 &17:01:08.914 & 64:16:12.554 & 2.214 & 24.55$\pm$0.16 & 0.07$\pm$0.08 & 0.57$\pm$0.11 & 3.14$\pm$0.36 & \nodata & 23.67$\pm$0.29 & \nodata & \nodata \\
BX1047 &17:00:23.369 & 64:15:31.444 & 2.231 & 24.55$\pm$0.16 & 0.43$\pm$0.11 & 1.04$\pm$0.16 & 4.01$\pm$0.16 & \nodata & 22.02$\pm$0.10 & \nodata & 22.61$\pm$0.31 \\
BX1084 &17:01:32.349 & 64:15:50.699 & 2.245 & 23.76$\pm$0.12 & 0.45$\pm$0.07 & 1.05$\pm$0.11 & 3.89$\pm$0.10 & \nodata & 22.14$\pm$0.46 & \nodata & \nodata \\
BX931 &17:00:25.497 & 64:14:29.805 & 2.246 & 25.15$\pm$0.16 & 0.47$\pm$0.12 & 1.12$\pm$0.28 & 3.60$\pm$0.28 & \nodata & 22.83$\pm$0.11 & \nodata & \nodata \\
BX794 &17:00:47.301 & 64:13:18.702 & 2.253 & 23.60$\pm$0.12 & 0.35$\pm$0.06 & 0.58$\pm$0.09 & 3.10$\pm$0.18 & \nodata & 23.67$\pm$0.25 & \nodata & \nodata \\
BX1030 &17:00:46.895 & 64:15:23.275 & 2.285 & 25.11$\pm$0.16 & 0.47$\pm$0.12 & 1.17$\pm$0.28 & 4.02$\pm$0.22 & \nodata & 22.32$\pm$0.18 & \nodata & \nodata \\
BX709 &17:00:54.761 & 64:12:19.973 & 2.285 & 25.18$\pm$0.16 & 0.05$\pm$0.10 & 0.77$\pm$0.15 & 3.06$\pm$0.59 & 23.37$\pm$0.38 & 23.06$\pm$0.28 & \nodata & \nodata \\
BX898 &17:01:05.284 & 64:14:06.570 & 2.286 & 24.48$\pm$0.14 & 0.18$\pm$0.06 & 0.57$\pm$0.11 & 2.37$\pm$0.51 & \nodata & 22.17$\pm$0.21 & \nodata & \nodata \\
MD69  & 17:00:47.610 & 64:09:44.78 & 2.289 & 24.85$\pm$0.16 & 0.37$\pm$0.09 & 1.50$\pm$0.33 & 4.87$\pm$0.12 & 21.13$\pm0.03$ &  20.89$\pm$0.04 & 20.59$\pm$0.07 & 20.63$\pm$0.11 \\
BX810 &17:01:31.117 & 64:13:29.357 & 2.290 & 24.68$\pm$0.16 & 0.36$\pm$0.09 & 0.87$\pm$0.18 & 3.99$\pm$0.19 & 22.92$\pm$0.23 & 22.78$\pm$0.23 & \nodata & \nodata \\
MD157 &17:00:52.191 & 64:15:29.245 & 2.293 & 24.35$\pm$0.14 & 0.35$\pm$0.07 & 1.65$\pm$0.14 & 4.46$\pm$0.10 & \nodata & 21.78$\pm$0.10 & \nodata & 21.57$\pm$0.12 \\
MD109 &17:01:04.482 & 64:12:09.293 & 2.294 & 25.46$\pm$0.16 & 0.26$\pm$0.10 & 1.44$\pm$0.37 & 3.60$\pm$0.28 & 23.12$\pm$0.11 & 22.86$\pm$0.10 & \nodata & \nodata \\
BX563 &17:01:15.755 & 64:10:27.088 & 2.296 & 23.82$\pm$0.12 & 0.42$\pm$0.07 & 0.86$\pm$0.09 & 3.15$\pm$0.18 & \nodata & 21.80$\pm$0.28 & \nodata & \nodata \\
BX939 &17:00:52.915 & 64:14:36.023 & 2.296 & 24.46$\pm$0.14 & 0.38$\pm$0.07 & 0.69$\pm$0.13 & 3.46$\pm$0.19 & \nodata & 23.05$\pm$0.13 & \nodata & \nodata \\
BX984 &17:01:02.164 & 64:14:56.440 & 2.299 & 23.51$\pm$0.12 & 0.17$\pm$0.04 & 0.72$\pm$0.06 & 3.44$\pm$0.14 & \nodata & 22.23$\pm$0.10 & \nodata & \nodata \\
BX919 &17:00:33.909 & 64:14:21.626 & 2.301 & 24.43$\pm$0.14 & 0.03$\pm$0.06 & 0.65$\pm$0.10 & 2.70$\pm$0.41 & \nodata & 23.51$\pm$0.23 & \nodata & \nodata \\
BX917 &17:01:16.110 & 64:14:19.801 & 2.302 & 24.43$\pm$0.14 & 0.28$\pm$0.07 & 0.95$\pm$0.13 & 4.04$\pm$0.12 & \nodata & 22.05$\pm$0.13 & \nodata & 22.19$\pm$0.37 \\
BX1133 &17:01:03.294 & 64:16:19.717 & 2.303 & 24.38$\pm$0.14 & 0.14$\pm$0.06 & 0.82$\pm$0.13 & 3.33$\pm$0.23 & \nodata & 22.85$\pm$0.10 & \nodata & \nodata \\
BX879 &17:01:17.768 & 64:13:55.116 & 2.308 & 23.50$\pm$0.12 & 0.61$\pm$0.04 & 1.40$\pm$0.19 & 2.84$\pm$0.22 & 22.22$\pm$0.34 & 22.22$\pm$0.15 & \nodata & \nodata \\
BX505 &17:00:48.224 & 64:10:05.862 & 2.309 & 25.17$\pm$0.16 & 0.45$\pm$0.12 & 1.28$\pm$0.26 & 4.36$\pm$0.19 & 22.23$\pm$0.10 & 22.11$\pm$0.10 & 22.49$\pm$0.17 & 22.10$\pm$0.15 \\
BX585 &17:01:13.491 & 64:10:58.829 & 2.311 & 24.71$\pm$0.16 & 0.12$\pm$0.08 & 0.76$\pm$0.13 & 3.15$\pm$0.36 & \nodata & \nodata & 23.56$\pm$0.41 & \nodata \\
MD103 &17:01:00.209 & 64:11:55.576 & 2.315 & 24.23$\pm$0.14 & 0.46$\pm$0.09 & 1.49$\pm$0.27 & 3.97$\pm$0.12 & 21.32$\pm$0.10 & 21.23$\pm$0.10 & 21.34$\pm$0.14 & \nodata \\
BX532 &17:00:40.154 & 64:10:22.537 & 2.331 & 23.76$\pm$0.12 & 0.22$\pm$0.06 & 0.57$\pm$0.09 & 3.63$\pm$0.12 & 21.02$\pm$0.10 & 21.10$\pm$0.10 & 21.21$\pm$0.16 & 21.56$\pm$0.21 \\
MD94  &17:00:42.020 & 64:11:24.224 & 2.333 & 24.72$\pm0.16$ & 0.94$\pm0.15$ & 2.06$\pm0.25$ & 5.15$\pm$0.15 & 20.59$\pm0.08$ & 20.43$\pm$0.03  & 20.35$\pm$0.22 & 20.28$\pm$0.14 \\
BX772 &17:00:53.076 & 64:13:04.566 & 2.346 & 24.96$\pm$0.16 & 0.20$\pm$0.09 & 0.57$\pm$0.14 & 3.29$\pm$0.31 & \nodata & 23.72$\pm$0.25 & \nodata & \nodata \\
MD174 &17:00:54.542 & 64:16:24.760 & 2.347 & 24.56$\pm$0.16 & 0.32$\pm$0.09 & 1.50$\pm$0.27 & 4.74$\pm$0.11 & \nodata & 20.51$\pm$0.10 & \nodata & $18.51\pm0.10$ \\
BX490 &17:01:14.831 & 64:09:51.691 & 2.403 & 22.88$\pm$0.06 & 0.36$\pm$0.03 & 0.92$\pm$0.05 & 2.88$\pm$0.10 & 22.05$\pm$0.10 & 22.08$\pm$0.10 & 21.95$\pm$0.11 & 22.13$\pm$0.30 \\
BX581 &17:01:02.726 & 64:10:51.299 & 2.406 & 23.87$\pm$0.12 & 0.28$\pm$0.06 & 0.62$\pm$0.10 & 3.21$\pm$0.15 & 23.11$\pm$0.15 & 23.17$\pm$0.14 & 22.60$\pm$0.21 & 21.94$\pm$0.19 \\
BX759 &17:00:59.545 & 64:12:55.455 & 2.418 & 24.43$\pm$0.14 & 0.36$\pm$0.07 & 1.29$\pm$0.18 & 3.24$\pm$0.23 & 21.81$\pm$0.31 & 22.39$\pm$0.29 & \nodata & \nodata \\
BX561 &17:01:04.180 & 64:10:43.834 & 2.426 & 24.65$\pm$0.16 & 0.19$\pm$0.08 & 1.04$\pm$0.16 & 4.63$\pm$0.11 & 21.54$\pm$0.10 & 21.35$\pm$0.10 & 21.19$\pm$0.10 & 20.85$\pm$0.10 \\
BX575 &17:01:03.341 & 64:10:50.932 & 2.431 & 23.82$\pm$0.12 & 0.16$\pm$0.04 & 0.74$\pm$0.06 & 2.60$\pm$0.26 & 23.10$\pm$0.12 & 23.00$\pm$0.10 & 23.03$\pm$0.41 & \nodata \\
BX717 &17:00:56.995 & 64:12:23.763 & 2.438 & 24.78$\pm$0.16 & 0.34$\pm$0.09 & 0.47$\pm$0.13 & 2.73$\pm$0.54 & \nodata & 23.89$\pm$0.24 & \nodata & \nodata \\
BX826 &17:01:17.163 & 64:13:41.960 & 2.438 & 24.45$\pm$0.14 & 0.02$\pm$0.06 & 0.52$\pm$0.09 & 2.94$\pm$0.28 & 22.65$\pm$0.25 & 22.55$\pm$0.10 & \nodata & \nodata \\
BX523 &17:00:41.712 & 64:10:14.884 & 2.471 & 24.51$\pm$0.16 & 0.46$\pm$0.11 & 1.28$\pm$0.18 & 3.52$\pm$0.25 & 22.74$\pm$0.20 & 22.68$\pm$0.15 & \nodata & 22.07$\pm$0.35 \\
BX553 &17:01:02.200 & 64:10:39.390 & 2.474 & 24.49$\pm$0.14 & 0.14$\pm$0.06 & 0.44$\pm$0.11 & 2.77$\pm$0.41 & 23.13$\pm$0.13 & 23.18$\pm$0.19 & \nodata & \nodata \\
MD72 &17:00:42.777 & 64:10:05.346 & 2.533 & 24.63$\pm$0.16 & 0.45$\pm$0.11 & 1.51$\pm$0.00 & 3.42$\pm$0.25 & 22.44$\pm$0.15 & 22.30$\pm$0.20 & \nodata & \nodata \\
BX568 &17:01:28.789 & 64:10:46.682 & 2.537 & 23.93$\pm$0.12 & 0.38$\pm$0.06 & 0.94$\pm$0.09 & 3.20$\pm$0.15 & 22.75$\pm$0.10 & 22.81$\pm$0.10 & 22.82$\pm$0.23 & 22.87$\pm$0.25 \\
BX588 &17:01:01.469 & 64:11:00.301 & 2.539 & 24.67$\pm$0.16 & 0.53$\pm$0.11 & 1.17$\pm$0.19 & 3.22$\pm$0.31 & 24.22$\pm$0.29 & \nodata & \nodata & 22.97$\pm$0.29 \\
BX609 &17:01:9.988 & 64:11:15.663 & 2.573 & 24.12$\pm$0.14 & 0.56$\pm$0.09 & 1.33$\pm$0.18 & 3.06$\pm$0.27 & 23.10$\pm$0.12 & 22.94$\pm$0.13 & 23.58$\pm$0.46 & 23.16$\pm$0.38 \\
MD154 &17:01:38.387 & 64:14:57.372 & 2.629 & 23.23$\pm$0.11 & 0.73$\pm$0.04 & 1.91$\pm$0.22 & 3.55$\pm$0.09 & 21.50$\pm$0.27 & \nodata & \nodata & \nodata \\
BX591 &17:00:50.179 & 64:11:01.162 & 2.686 & 24.68$\pm$0.16 & 0.22$\pm$0.09 & 1.06$\pm$0.16 & 2.93$\pm$0.54 & 23.48$\pm$0.14 & 23.91$\pm$0.23 & \nodata & \nodata \\
MD128 &17:00:30.292 & 64:13:15.165 & 2.713 & 23.41$\pm$0.11 & 0.29$\pm$0.04 & 1.24$\pm$0.08 & 2.21$\pm$0.28 & \nodata & 23.67$\pm$0.16 & \nodata & \nodata \\
MD98 &17:01:04.387 & 64:11:49.543 & 2.738 & 24.54$\pm$0.16 & 0.70$\pm$0.10 & 1.82$\pm$0.46 & 2.94$\pm$0.54 & 22.62$\pm$0.10 & 22.94$\pm$0.22 & \nodata & \nodata \\
BX928 &17:01:19.623 & 64:14:25.911 & 2.755 & 23.77$\pm$0.12 & 0.66$\pm$0.07 & 1.54$\pm$0.19 & 2.95$\pm$0.22 & \nodata & 22.99$\pm$0.21 & \nodata & \nodata \\
MD152 &17:00:31.085 & 64:14:54.048 & 2.791 & 23.43$\pm$0.11 & 0.86$\pm$0.05 & 1.92$\pm$0.29 & 2.92$\pm$0.15 & \nodata & 21.62$\pm$0.10 & \nodata & 21.65$\pm$0.11 \\
BX529 &17:00:44.618 & 64:10:19.001 & 2.875 & 24.27$\pm$0.14 & 0.25$\pm$0.06 & 0.28$\pm$0.06 & 2.74$\pm$0.41 & 23.35$\pm$0.25 & \nodata & \nodata & \nodata \\
MD155 &17:00:47.888 & 64:15:05.446 & 2.875 & 24.03$\pm$0.14 & 0.56$\pm$0.09 & 1.83$\pm$0.43 & 2.32$\pm$0.51 & \nodata & 23.31$\pm$0.14 & \nodata & \nodata \\
MD126 &17:00:51.610 & 64:13:05.346 & 2.898 & 23.94$\pm$0.12 & 0.67$\pm$0.04 & 2.01$\pm$0.46 & 2.32$\pm$0.42 & \nodata & 22.82$\pm$0.15 & \nodata & \nodata \\
\enddata
\tablenotetext{a}{Includes all galaxies with spectroscopic redshifts that are significantly detected in the ground-based 
$K_s$-band image and in one or more of the IRAC bands.}
\tablenotetext {b}{${\cal R}(AB)-K_s$(Vega), to make the comparison with other samples easier. This color can be converted
to the AB system used in the fitting and in the plots in Figure~\ref{fig:seds1} \\ by subtracting 1.82.}
\end{deluxetable}
\clearpage
\end{landscape}

\clearpage
\LongTables
\begin{landscape}
\begin{deluxetable}{lcccccccccc}
%\rotate
\tabletypesize{\footnotesize}
\tablewidth{0pt}
\tablecaption{Modeling Results\label{tab:mod}}
\tablehead{
\colhead{Object} &  \colhead{z} & \colhead{log M$_{\ast}$(CSF)\tablenotemark{a}} & \colhead{E(B-V)} & \colhead{SFR\tablenotemark{b}} &
\colhead{Age\tablenotemark{c}} & \colhead{log M$_{\ast}$($\tau$)\tablenotemark{d}} & \colhead{E(B-V)} &
\colhead{SFR\tablenotemark{e}} & \colhead{Age\tablenotemark{f}} & \colhead{$\tau$\tablenotemark{g}} }
\startdata
BX627 & 1.478 & ~9.42$\pm$0.36 & 0.05 & ~~~7 & ~404 & ~9.42$\pm$0.43 & 0.05 & ~~~7 & ~404 & const \\
BX738 & 1.710 & 10.40$\pm$0.09 & 0.10 & ~~20 & 1278 & 10.35$\pm$0.05 & 0.00 & ~~~4 & ~404 & ~100 \\
BX756 & 1.738 & ~9.76$\pm$0.05 & 0.34 & ~377 & ~~15 & ~9.76$\pm$0.05 & 0.34 & ~377 & ~~15 & const \\
BX681 & 1.740 & 10.50$\pm$0.24 & 0.21 & ~246 & ~128 & 10.65$\pm$0.03 & 0.08 & ~~~1 & ~~81 & ~~10 \\
BX843 & 1.771 & ~9.89$\pm$0.23 & 0.21 & ~~24 & ~321 & 10.06$\pm$0.14 & 0.04 & ~~~1 & ~255 & ~~50 \\
BX470 & 1.841 & 10.81$\pm$0.09 & 0.30 & ~~64 & 1015 & 10.75$\pm$0.07 & 0.23 & ~~24 & ~321 & ~100 \\
BX846 & 1.843 & 10.44$\pm$0.10 & 0.12 & ~~27 & 1015 & 10.39$\pm$0.08 & 0.03 & ~~~7 & ~360 & ~100 \\
BX574 & 1.846 & 10.16$\pm$0.19 & 0.24 & ~~51 & ~286 & 10.34$\pm$0.10 & 0.06 & ~~~3 & ~255 & ~~50 \\
BX1043 & 1.847 & ~9.91$\pm$0.25 & 0.10 & ~~32 & ~255 & ~9.94$\pm$0.10 & 0.00 & ~~~0 & ~~90 & ~~10 \\
BX1087 & 1.871 & 10.24$\pm$0.11 & 0.15 & ~~68 & ~255 & 10.31$\pm$0.08 & 0.04 & ~~~3 & ~114 & ~~20 \\
BX1032 & 1.883 & 10.08$\pm$0.16 & 0.20 & ~~53 & ~227 & 10.12$\pm$0.11 & 0.10 & ~~~4 & ~102 & ~~20 \\
BX782 & 1.930 & 10.39$\pm$0.11 & 0.10 & ~~24 & 1015 & 10.33$\pm$0.07 & 0.00 & ~~~6 & ~360 & ~100 \\
BX530 & 1.942 & ~9.84$\pm$0.21 & 0.34 & ~699 & ~~10 & 10.29$\pm$0.09 & 0.08 & ~~~1 & ~~72 & ~~10 \\
BX807 & 1.971 & ~9.86$\pm$0.16 & 0.10 & ~~10 & ~719 & ~9.85$\pm$0.14 & 0.01 & ~~~3 & ~321 & ~100 \\
BX903 & 1.976 & 10.10$\pm$0.16 & 0.14 & ~~22 & ~571 & 10.11$\pm$0.20 & 0.10 & ~~13 & ~360 & ~200 \\
BX536 & 1.977 & 10.65$\pm$0.06 & 0.23 & ~138 & ~321 & 10.74$\pm$0.06 & 0.09 & ~~20 & ~203 & ~~50 \\
BX412 & 1.981 & 10.24$\pm$0.18 & 0.20 & ~~61 & ~286 & 10.28$\pm$0.15 & 0.12 & ~~16 & ~161 & ~~50 \\
BX526 & 2.018 & 10.06$\pm$0.15 & 0.04 & ~~16 & ~719 & ~9.97$\pm$0.18 & 0.00 & ~~~8 & ~255 & ~100 \\
BX1075 & 2.061 & ~8.98$\pm$0.31 & 0.30 & ~115 & ~~~8 & ~8.97$\pm$0.31 & 0.30 & ~~88 & ~~~7 & ~~10 \\
BX625 & 2.078 & 10.04$\pm$0.12 & 0.07 & ~~12 & ~905 & 10.02$\pm$0.10 & 0.01 & ~~~6 & ~454 & ~200 \\
BX592 & 2.099 & ~9.30$\pm$0.22 & 0.37 & ~199 & ~~10 & ~9.82$\pm$0.24 & 0.12 & ~~~2 & ~102 & ~~20 \\
BX535 & 2.113 & ~9.48$\pm$0.24 & 0.43 & ~199 & ~~15 & 10.00$\pm$0.29 & 0.23 & ~~~1 & ~~72 & ~~10 \\
BX1012 & 2.115 & 10.10$\pm$0.20 & 0.07 & ~~25 & ~509 & 10.06$\pm$0.19 & 0.01 & ~~10 & ~255 & ~100 \\
BX691 & 2.190 & 11.04$\pm$0.03 & 0.29 & ~~40 & 2750 & 11.14$\pm$0.05 & 0.13 & ~~~9 & 2750 & 1000 \\
BX1118 & 2.214 & ~9.97$\pm$0.20 & 0.07 & ~~13 & ~719 & ~9.86$\pm$0.20 & 0.00 & ~~~4 & ~286 & ~100 \\
BX1047 & 2.231 & 10.71$\pm$0.12 & 0.26 & ~~50 & 1015 & 10.67$\pm$0.07 & 0.12 & ~~~8 & ~404 & ~100 \\
BX1084 & 2.245 & 11.09$\pm$0.11 & 0.26 & ~120 & 1015 & 10.92$\pm$0.09 & 0.10 & ~~15 & ~404 & ~100 \\
BX931 & 2.246 & 10.19$\pm$0.27 & 0.27 & ~~38 & ~404 & 10.24$\pm$0.27 & 0.20 & ~~15 & ~255 & ~100 \\
BX794 & 2.253 & ~9.49$\pm$0.36 & 0.29 & ~430 & ~~~7 & ~9.45$\pm$0.41 & 0.28 & ~265 & ~~~7 & ~~10 \\
BX1030 & 2.285 & 10.65$\pm$0.17 & 0.27 & ~~39 & 1139 & 10.58$\pm$0.17 & 0.23 & ~~22 & ~454 & ~200 \\
BX709 & 2.285 & 10.37$\pm$0.17 & 0.10 & ~~10 & 2400 & 10.37$\pm$0.27 & 0.10 & ~~10 & 2400 & const \\
BX898 & 2.286 & 10.73$\pm$0.11 & 0.10 & ~~20 & 2750 & 10.73$\pm$0.21 & 0.10 & ~~20 & 2750 & const \\
MD69 & 2.289 & 11.46$\pm$0.02 & 0.35 & ~105 & 2750 & 11.51$\pm$0.05 & 0.24 & ~~40 & 2200 & 1000 \\
BX810 & 2.290 & 10.51$\pm$0.12 & 0.20 & ~~29 & 1139 & 10.41$\pm$0.09 & 0.01 & ~~~3 & ~454 & ~100 \\
MD157 & 2.293 & 11.03$\pm$0.07 & 0.28 & ~~66 & 1609 & 10.89$\pm$0.07 & 0.21 & ~~22 & ~360 & ~100 \\
MD109 & 2.294 & 10.48$\pm$0.17 & 0.17 & ~~14 & 2100 & 10.48$\pm$0.25 & 0.17 & ~~14 & 2100 & const \\
BX563 & 2.296 & 10.57$\pm$0.23 & 0.21 & ~~92 & ~404 & 10.57$\pm$0.28 & 0.21 & ~~92 & ~404 & const \\
BX939 & 2.296 & 10.23$\pm$0.15 & 0.16 & ~~27 & ~640 & 10.28$\pm$0.11 & 0.00 & ~~~3 & ~404 & ~100 \\
BX984 & 2.299 & 10.65$\pm$0.07 & 0.12 & ~~44 & 1015 & 10.55$\pm$0.05 & 0.01 & ~~10 & ~360 & ~100 \\
BX919 & 2.301 & ~9.84$\pm$0.21 & 0.07 & ~~17 & ~404 & ~9.74$\pm$0.26 & 0.00 & ~~~5 & ~161 & ~~50 \\
BX917 & 2.302 & 10.96$\pm$0.07 & 0.18 & ~~33 & 2750 & 10.86$\pm$0.09 & 0.14 & ~~22 & 1015 & ~500 \\
BX1133 & 2.303 & 10.37$\pm$0.12 & 0.10 & ~~20 & 1139 & 10.27$\pm$0.11 & 0.01 & ~~~5 & ~360 & ~100 \\
BX879 & 2.308 & 10.03$\pm$0.10 & 0.42 & 1473 & ~~~7 & 10.02$\pm$0.13 & 0.41 & 1380 & ~~~7 & ~100 \\
BX505 & 2.309 & 10.79$\pm$0.13 & 0.28 & ~~38 & 1609 & 10.73$\pm$0.08 & 0.20 & ~~16 & ~571 & ~200 \\
BX585 & 2.311 & 10.03$\pm$0.28 & 0.11 & ~~17 & ~640 & ~9.99$\pm$0.26 & 0.01 & ~~~4 & ~321 & ~100 \\
MD103 & 2.315 & 11.07$\pm$0.13 & 0.28 & ~103 & 1139 & 10.96$\pm$0.13 & 0.21 & ~~38 & ~321 & ~100 \\
BX532 & 2.331 & 11.26$\pm$0.03 & 0.14 & ~~65 & 2750 & 11.33$\pm$0.08 & 0.09 & ~~36 & 2750 & 2000 \\
MD94 & 2.333 & 11.52$\pm$0.10 & 0.49 & ~324 & 1015 & 11.43$\pm$0.10 & 0.42 & ~113 & ~321 & ~100 \\
BX772 & 2.346 & 10.07$\pm$0.18 & 0.10 & ~~12 & 1015 & 10.06$\pm$0.16 & 0.03 & ~~~5 & ~509 & ~200 \\
MD174 & 2.347 & 11.59$\pm$0.04 & 0.34 & ~140 & 2750 & 11.65$\pm$0.05 & 0.20 & ~~40 & 2500 & 1000 \\
BX490 & 2.403 & ~9.92$\pm$0.12 & 0.29 & ~835 & ~~10 & ~9.88$\pm$0.33 & 0.28 & ~506 & ~~~9 & ~~10 \\
BX581 & 2.406 & 10.37$\pm$0.11 & 0.11 & ~~33 & ~719 & 10.36$\pm$0.10 & 0.10 & ~~26 & ~509 & ~500 \\
BX759 & 2.418 & 10.52$\pm$0.22 & 0.23 & ~~73 & ~454 & 10.52$\pm$0.28 & 0.23 & ~~73 & ~454 & const \\
BX561 & 2.426 & 11.31$\pm$0.02 & 0.29 & ~~79 & 2600 & 11.39$\pm$0.06 & 0.16 & ~~25 & 2400 & 1000 \\
BX575 & 2.431 & ~9.95$\pm$0.20 & 0.11 & ~~44 & ~203 & ~9.99$\pm$0.25 & 0.07 & ~~25 & ~161 & ~100 \\
BX717 & 2.438 & ~9.85$\pm$0.38 & 0.09 & ~~14 & ~509 & ~9.99$\pm$0.42 & 0.01 & ~~~6 & ~454 & ~200 \\
BX826 & 2.438 & 10.63$\pm$0.04 & 0.04 & ~~17 & 2600 & 10.69$\pm$0.06 & 0.01 & ~~11 & 2300 & 2000 \\
BX523 & 2.471 & 10.31$\pm$0.30 & 0.25 & ~~71 & ~286 & 10.39$\pm$0.28 & 0.18 & ~~28 & ~227 & ~100 \\
BX553 & 2.474 & 10.34$\pm$0.10 & 0.04 & ~~14 & 1609 & 10.30$\pm$0.10 & 0.01 & ~~10 & ~806 & ~500 \\
MD72 & 2.533 & 10.61$\pm$0.28 & 0.22 & ~~56 & ~719 & 10.61$\pm$0.30 & 0.22 & ~~56 & ~719 & const \\
BX568 & 2.537 & 10.27$\pm$0.27 & 0.16 & ~~52 & ~360 & 10.33$\pm$0.06 & 0.01 & ~~~5 & ~227 & ~~50 \\
BX588 & 2.539 & ~9.47$\pm$0.22 & 0.35 & ~295 & ~~10 & ~9.44$\pm$0.27 & 0.34 & ~210 & ~~10 & ~~20 \\
BX609 & 2.573 & ~9.68$\pm$0.15 & 0.35 & ~576 & ~~~8 & 10.17$\pm$0.25 & 0.00 & ~~~0 & ~128 & ~~10 \\
MD154 & 2.629 & 10.94$\pm$0.24 & 0.35 & ~677 & ~128 & 10.96$\pm$0.09 & 0.01 & ~~~6 & ~286 & ~~50 \\
BX591 & 2.686 & 10.10$\pm$0.19 & 0.06 & ~~16 & ~806 & 10.09$\pm$0.16 & 0.00 & ~~~7 & ~454 & ~200 \\
MD128 & 2.713 & ~9.38$\pm$0.09 & 0.18 & ~287 & ~~~8 & ~9.37$\pm$0.11 & 0.18 & ~223 & ~~~7 & ~~10 \\
MD98 & 2.738 & ~9.82$\pm$0.26 & 0.36 & ~436 & ~~15 & 10.47$\pm$0.27 & 0.07 & ~~~6 & ~227 & ~~50 \\
BX928 & 2.755 & ~9.85$\pm$0.13 & 0.35 & ~972 & ~~~7 & ~9.87$\pm$0.21 & 0.36 & ~895 & ~~~6 & ~~10 \\
MD152 & 2.791 & 10.28$\pm$0.06 & 0.41 & 2087 & ~~~9 & 10.28$\pm$0.16 & 0.41 & 2087 & ~~~9 & const \\
BX529 & 2.875 & 10.33$\pm$0.20 & 0.03 & ~~19 & 1139 & 10.28$\pm$0.24 & 0.01 & ~~15 & ~640 & ~500 \\
MD155 & 2.875 & ~9.52$\pm$0.18 & 0.26 & ~333 & ~~10 & ~9.48$\pm$0.25 & 0.24 & ~202 & ~~~9 & ~~10 \\
MD126 & 2.898 & ~9.72$\pm$0.09 & 0.29 & ~520 & ~~10 & ~9.69$\pm$0.23 & 0.28 & ~380 & ~~10 & ~~20 \\
\enddata
\tablenotetext{a}{Stellar mass and uncertainty, in solar units, for constant star formation model.}
\tablenotetext{b}{Star formation rate, in M$_{\sun}$ yr$^{-1}$, for constant star formation model.}
\tablenotetext{c}{Age, in Myr, for constant star formation model.}
\tablenotetext{d}{Stellar mass and uncertainty, in solar units, for the best-fitting $\tau$ model. 
The stellar mass uncertainty reflects the uncertainty
in $\tau$.}
\tablenotetext{e}{Star formation rate, in M$_{sun}$ yr$^{-1}$, for the best-fitting $\tau$ model.}
\tablenotetext{f}{Age, in Myr, for the best-fitting $\tau$ model.}
\tablenotetext{g}{Best-fitting star-formation decay timescale (SFR $\propto {\rm e}^{-\tau}$), in Myr. }
\end{deluxetable}
\clearpage
\end{landscape}

\begin{deluxetable}{lccccc}
%\tabletypesize{\scriptsize}
\tablewidth{0pt}
\tablecaption{Stellar Mass Estimates as a Function of $K_s$ Magnitude\label{tab:MvsK}}
\tablehead{
\colhead{Sample} & \colhead{$\langle M_{\ast}(\tau)\rangle$\tablenotemark{a}} & \colhead{$\langle M_{\ast}(CSF)\rangle$\tablenotemark{b}} 
& \colhead{$\langle M_{\ast}(\tau)\rangle$(no IRAC)\tablenotemark{c}} } 
\startdata
All (72) & $10.32\pm0.51$ & $10.30\pm 0.53$ & $10.37\pm0.55$ \\ 
$K_s<21$ (38) & $10.56\pm0.50$ & $10.55\pm0.51$ & $10.67\pm0.44$ \\ 
$K_s<20.6$ (20) & $10.75\pm0.55$ & $10.75\pm0.56$ & $10.88\pm0.40$ \\
$K_s<20$ (9) & $10.96\pm0.55$ & $10.96\pm0.55$ &  $11.01\pm0.40$ \\
\enddata
\tablenotetext{a}{Mean and standard deviation of the log of the inferred stellar mass for the best-fit $\tau$ models.}
\tablenotetext{b}{Mean and standard deviation of the log of the inferred stellar mass for the CSF models.}
\tablenotetext{c}{Mean and standard deviation of the log of the inferred stellar mass for the best-fit $\tau$ model,
excluding the IRAC photometry.}
\end{deluxetable}
\clearpage

\end{document}